\documentclass[preprintnumbers,article,amsmath,amssymb,floatfix,9pt,prd,onecolumn,
superscriptaddress,nofootinbib,amsfonts,showpacs]{revtex4-2}

\usepackage{bbm}
\usepackage{amsfonts}
\usepackage{mathrsfs}
\usepackage{latexsym}
\usepackage{epsfig}
\usepackage{epstopdf}
\usepackage{epstopdf}
\usepackage{graphicx}
\usepackage{amssymb}
\usepackage{amsmath}
\usepackage{dcolumn}
\usepackage{bm}
\usepackage{color}
\usepackage{comment}
\usepackage{xcolor}
\def\be{\begin{equation}}
\def\ee{\end{equation}}
\def\beq{\begin{eqnarray}}
\def\eeq{\end{eqnarray}}

\usepackage[latin1]{inputenc}
\usepackage{graphicx, psfrag}
\usepackage[colorlinks=true, citecolor=green, urlcolor = magenta, linkcolor= blue, bookmarks=true]{hyperref}
\usepackage{float}
\usepackage{amsfonts}
\usepackage{hyperref}
\usepackage{subfigure}
\usepackage{booktabs}
\usepackage{orcidlink}
\usepackage{enumitem}

\begin{document}
\title{Unraveling the mysteries of wormhole formation in Rastall-Rainbow gravity: A comprehensive study using the embedding approach} 

\author{Abdelghani Errehymy\orcidlink{0000-0002-0253-3578}}%
\email[Email:]{abdelghani.errehymy@gmail.com}
\affiliation{Astrophysics Research Centre, School of Mathematics, Statistics and Computer Science, University of KwaZulu-Natal, Private Bag X54001, Durban 4000, South Africa}

\author{Ayan Banerjee} %\orcidlink{0000-0003-3422-8233}} 
\email[Email:]{ayanbanerjeemath@gmail.com}
\affiliation{Astrophysics and Cosmology Research Unit, School of Mathematics, Statistics and Computer Science, University of KwaZulu--Natal, Private Bag X54001, Durban 4000, South Africa}

\author{Orhan Donmez}%\orcidlink{}}%
\email[Email:]{orhan.donmez@aum.edu.kw} 
\affiliation{College of Engineering and Technology, American University of the Middle East, Egaila 54200, Kuwait}

\author{Mohammed Daoud}%\orcidlink{0000-0002-8494-2796}}% 
\email[Email:]{m$_{}$daoud@hotmail.com}
\affiliation{Department of Physics, Faculty of Sciences, Ibn Tofail University, P.O. Box 133, Kenitra 14000, Morocco}
\affiliation{Abdus Salam International Centre for Theoretical Physics, Miramare, Trieste 34151, Italy}

\author{Kottakkaran Sooppy Nisar}%\orcidlink{}}%
\email[Email:]{n.sooppy@psau.edu.sa}
\affiliation{Department of Mathematics, College of Science and Humanities in Alkharj, Prince Sattam bin Abdulaziz University, Al Kharj, Saudi Arabia}

%\author{Meraj Ali Khan}%\orcidlink{0000-0001-6554-1228}}%
%\email[Email:]{mskhan@imamu.edu.sa} 
%\affiliation{Department of Mathematics and Statistics, College of Science, Imam Mohammad Ibn Saud Islamic University (IMSIU), P.O. Box 65892, Riyadh 11566, Saudi Arabia}

\author{Abdel-Haleem Abdel-Aty}%\orcidlink{0000-0002-6763-2569}}%
\email[Email:]{amabdelaty@ub.edu.sa}
\affiliation{Department of Physics, College of Sciences, University of Bisha, P.O. Box 344, Bisha 61922, Saudi Arabia}

\begin{abstract}
The present work looks for the possible existence of static and spherically symmetric wormhole geometries in Rastall-Rainbow gravity. Since, the Rastall-Rainbow gravity model has been constructed with the combination of Rastall theory and the gravity's rainbow formalism. Taking advantage of the Karmarkar condition for embedding class one metrics, we solve the modified field equations analytically that describe wormholes for specific choice of redshift function. For specific parameter ranges, the solution represents a traversable wormhole that exhibits the violation of null energy condition and consequently the weak energy condition also. Furthermore, we focus on the wormhole stability via adiabatic sound velocity analysis. This model establishes a strong connection between two model parameters, namely, the Rastall parameters and the Rainbow functions, and how it affects the wormhole solution.

\end{abstract}

\maketitle

\date{\today}

\section{Introduction}\label{Sec1}

Traversable Lorentzian wormholes \cite{Morris:1988cz} are of fascinating objects predicted by Einstein's theory of gravity. To be specific, wormholes are handle-like structures that connect two spatially separated locations in our universe or two different universes. Historically, the idea of wormholes was first theorized by Flamm in 1916 \cite{Flamm:1916}, shortly afterwards Einstein proposed his general relativity (GR) theory. Almost twenty years later, Albert Einstein and Nathan Rosen \cite{Einstein:1953tkd}
proposed the existence of ``bridges" between two different points in spacetime. This idea open up the possibility of a shorter route that
could reduce travel time and distance also. This simplest form of mathematical model was dubbed ``Einstein-Rosen bridges" for what the physicist John Wheeler would later coin a word ``wormhole." However, their solution is prohibited by physical rules due to the presence of a physical singularity. But, it was Morris and Thorne \cite{Morris:1988cz}, who came up with an astonishing idea that could be a two-way passage through them without processing any event horizon. According to them, a traversable wormhole has a throat that connects two asymptotically flat regions. In this setting wormholes throat are identified as a minimal surface area satisfying the flare-out condition \cite{Morris:1988cz}. On the other hand Morris, Thorne and Yurtsever \cite{Morris1988} studied the stability of transforming wormholes into a time machine. Visser systematically systematized all these seminal works in his book ``Lorentzian wormholes: from Einstein to Hawking" \cite{Visser:1995}.

The fundamental problem associated with traversable wormholes is the presence of matter that violates the energy conditions
to satisfy the flare-out condition \cite{Morris1988}. Such a matter is referred to as ``exotic matter" (the stress-energy tensor of matter that violates the null energy condition). The presence of exotic matter has a repulsive nature that prevents the throat from shrinking. However, Visser \textit{et al} \cite{Visser:2003yf} had pointed out that the matter violating the averaged null energy condition to support traversable wormholes can be made infinitesimally small by means of an appropriate choice of wormhole geometry. Further, the amount of exotic matter can be confined at the wormhole throat by the cut-and-paste technique, yielding a thin-shell wormhole solution can be found in Refs. \cite{Visser:1989kh,Visser:1989kg}.
Thin-shell wormholes have been studied by many authors \cite{Lobo:2003xd,Dias:2010uh,Eiroa:2007qz,Usmani:2010cd}.

These objects have been extensively studied within the context of GR. The Morris-Thorne wormholes
in the presence of a generic cosmological constant has been studied in \cite{Lemos:2003jb}. Wormhole solutions have also been obtained in support of phantom energy EoS \cite{Sushkov:2005kj,Lobo:2005us,Gonzalez:2009cy}.  The key argument is that the energy density of phantom field increases with time and thus provides a notion for the existence of wormholes. Lobo \cite{Lobo:2005yv} discussed the stability equilibrium configurations for the specific
phantom wormhole geometries by applying the linearized spherically symmetric perturbations. In the same line of action, various solutions regarding wormholes were found for generalized Chaplygin gas \cite{Lobo:2005vc,Kuhfittig:2009mx,Sharif:2014opa}, 
varying cosmological constant \cite{Rahaman:2006xa}, polytropic phantom energy \cite{Jamil:2010ziq}  and ghost scalar fields \cite{Carvente:2019gkd}. 
Subsequently, evolving wormholes have also been studied in support of phantom matter and cosmological constant 
\cite{Cataldo:2008ku,Cataldo:2011zn}, see also Refs. \cite{Setare:2015iqa,Bhattacharya:2021frx}.

Therefore, it goes without saying that the theoretical construction of wormhole geometries with ordinary matter is a challenging task in gravitation physics. But, we are still far away from the success-line at least in GR context. To overcome this situation stated above, one usually considers modified theories of gravity. Since, the modified theories introduce a new degree of freedom that helps a sustainable wormhole solution while keeping the matter components non-exotic. In alternative theories of gravity, for example, higher order gravity theories \cite{Hochberg:1990is,Ghoroku:1992tz}, higher-dimensional cosmological wormholes \cite{Zangeneh:2014noa} and in the Einstein-Gauss-Bonnet theory \cite{Bhawal:1992sz,Maeda:2008nz,Mehdizadeh:2015jra} authors have studied wormhole solutions. In support of this, one can find traversable wormholes in $f(R)$ gravity with ordinary matter \cite{Pavlovic:2014gba,Lobo:2009ip}, 
or sourced by dark matter \cite{Muniz:2022eex,Errehymy:2024yey}. Indeed, wormhole geometries have also been explored in distinct
modifications of GR, namely: Extended theories of gravity \cite{DeFalco:2021klh,DeFalco:2021ksd,Errehymy:2023rnd,Errehymy:2023rsm,Errehymy:2024spg,Mustafa:2024jsv,Errehymy:2024cgy}, third-order Lovelock gravity \cite{KordZangeneh:2015dks,Mehdizadeh:2016nna}, hybrid metric-Palatini gravity \cite{Rosa:2021yym,KordZangeneh:2020ixt} and $f(Q)$ gravity \cite{Banerjee:2021mqk,Parsaei:2022wnu,Hassan:2022hcb}. A similar analysis has been performed in $f(R,T)$ gravity, see
Refs. \cite{Moraes:2017mir,Elizalde:2018frj,Moraes:2019pao}. Interestingly, authors in \cite{Zubair:2019uul,Rosa:2022osy,Banerjee:2020uyi}
have shown that one can theoretically construct exact wormholes solutions without requiring exotic matter.

In \cite{Mota:2019zln}, authors have proposed a new modified gravity theory, namely, the Rastall-Rainbow gravity. Since, the Rastall-Rainbow gravity model has been constructed with the combination of Rastall theory \cite{Rastall:1972swe} and the gravity's rainbow formalism \cite{Magueijo:2002xx}. Thus, the closed-form wormhole solution in Rastall-Rainbow gravity has received significant attention from the researchers. The focal point of our discussion is the space-time of embedding class-one. A spacetime is said to be of embedding class $N$ if it can be locally embed in an $(4+N)$-dimensional pseudo-Riemannian spacetime. Interestingly, the idea of embedding establishes a geometric relation between the metric functions and their derivatives for general spherically symmetrical line-element \cite{Karmarkar}. This means that one can choose one of the metric functions and generate the other metric function. This idea, the so-called ``Karmarkar condition" 
has been used to address new solutions of relativistic astrophysical solutions, such as compact stars \cite{Singh:2020iqh,Nikolaev:2020ztr,Maurya:2021sju,Singh:2016xei} and in wormhole solutions \cite{Agrawal:2021gdq,Tello-Ortiz:2020zfs} also.

The above discussion motivated us to search for wormhole solutions within the Rastall-Rainbow gravity context. The 
main attraction is the use of Karmarkar conditions on wormhole geometries and investigate the effects of the Rastall-Rainbow parameters on it. Our work plan is:  In Section \ref{sec2}, we briefly review the Rastall-Rainbow gravity theory and derive the corresponding modified field equations for the configurations under consideration. In the same section we establish a relation between two metric functions via embedding approach. In Section \ref{Sec3}, we choose a fairly simple redshift function that ensures the space-time is free from horizons and singularities. Using the chosen redshift function, we derive the shape function 
and studied all the constraints for wormhole solution. The structure equation for standard energy conditions is presented in Section \ref{Sec4}. In Section \ref{Sec5}, we present some features concerning wormhole solution under consideration, specifically, adiabatic sound velocity and energy conditions. Finally, Section \ref{Sec6} is devoted to summarize and discuss our findings.

%-----------------------------------------------------
\section{Rastall-Rainbow gravity and wormholes}\label{sec2}

\subsection{Rastall-Rainbow theory}
\label{TOV_RR}
In Einstein's GR, the conservation law of the energy-momentum tensor is stated as $T_{\mu;\nu}^{\nu}=0$. Nevertheless, Rastall gravity is a modification of the GR suggested by Rastall \cite{Rastall:1972swe}. In Rastall gravity, the conservation law of the energy-momentum tensor is not conserved, and expressed in the following form
\begin{equation}\label{1}
T_{\mu;\nu}^{\nu}=\bar{\lambda}R_{,\mu}
\end{equation}
where $\bar{\lambda}=\frac{1-\lambda}{16\pi G}$. Here, $\lambda$ stands for the Rastall parameter, which is a constant denoting the deviation from GR and describing the matter field's affinity for coupling with geometry. When $\lambda=1$, the standard law of conservation is recovered. What's more, in the case of a flat space-time where the Ricci scalar $R=0$, the standard conservation law also holds. On the other hand, in the presence of Rastall gravity, where $\lambda\neq 1$ and space-time is non-flat, the above formula captures the modified conservation law effect. The aforementioned formula can be expressed as follows, 
\begin{equation}
\left(T_{\mu}^{\nu}-\bar{\lambda}\delta_{\mu}^{\nu}R\right)_{;\nu}=0.
\end{equation}
So, in the case of Rastall gravity, Einstein's equation can be modified to take a slightly different form, expressed as follows,
\begin{equation}
R_{\mu}^{\nu}-\frac{1}{2}\delta_{\mu}^{\nu}R=8\pi G
\left(T_{\mu}^{\nu}-\bar{\lambda}\delta_{\mu}^{\nu}R\right),
\end{equation}
which can be further simplified to the following form,
\begin{equation}\label{eq.4}
R_{\mu}^{\nu}-\frac{\lambda}{2}\delta_{\mu}^{\nu}R=8\pi G
T_{\mu}^{\nu}.
\end{equation}
The corresponding expression for the trace of the stress-energy tensor is now obtained in the following way, 
\begin{equation}\label{eq.5}
T=\frac{1-2\lambda}{8\pi G}R.
\end{equation}
Accordingly, the modified form of Einstein's equation in Rastall's gravity can be expressed as,
\begin{equation}
R_{\mu}^{\nu}-\frac{1}{2}\delta_{\mu}^{\nu}R=8\pi G
\left(T_{\mu}^{\nu}-\frac{1}{2}\frac{1-\lambda}{1-2\lambda}\delta_{\mu}^{\nu}T\right).
\end{equation}

In 2002, Magueijo and Smolin \cite{Magueijo:2001cr} introduced the gravitational rainbow as an extension of doubly special relativity to curved spacetimes. The gravitational rainbow involves a distortion of spacetime caused by two arbitrary functions $\Pi(x)$ and $\Sigma(x)$, known as rainbow functions.  These functions are governed by the following conditions:
\begin{equation}
{\cal E}^{2}\Pi^{2}(x)-\upsilon^{2}\Sigma^{2}(x)=m^{2}.
\end{equation}
Here, we are defining $x$ as the ratio between ${\cal E}$ and ${\cal E}_{Pl}$ i.e., $x={\cal E}/{\cal E}_{Pl}$. In this context, ${\cal E}$ stands for energy, $\upsilon$ represents momentum, $m$ denotes the mass of a particle under test, and ${\cal E}_{Pl}=\sqrt{\hbar c^{5}}/G$ corresponds to the Planck energy. Awad \textit{et al} \cite{Awad:2013nxa} and Khodadi \textit{et al} \cite{Khodadi:2016bcx} selected $\Pi(x)=1$ and $\Sigma(x)=\sqrt{1+x^{2}}$ for studying solutions associated with a nonsingular universe. In addition, the exponential form of the rainbow has been employed in \cite{Awad:2013nxa, Amelino-Camelia:1997ieq} for investigating gamma-ray bursts. In the absence of test particles, the rainbow functions fulfill the following conditions:
\begin{equation}
\lim_{x \to 0}\Pi(x)=1,~\lim_{x \to 0}\Sigma(x)=1.
\end{equation}

In this context, the metric characterizing space-time depends on the energy \cite{Magueijo:2002xx}, and can be expressed as,
\begin{equation}
g^{\mu\nu}(x)=\eta^{ab} e_{a}^{\mu}(x)\otimes e_{b}^{\nu}(x).
\label{eq3}
\end{equation}
In this case, the energy-dependent Vierbein fields $e_{a}^{\mu}(x)$ are connected to the independent Vierbein fields $\widetilde{e}_{a}^{\mu}$ by the following relationships:
\begin{equation}
e_{0}^{\mu}(x)=\frac{1}{\Pi(x)} \widetilde{e}_{0}^{\mu}, \quad e_{k}^{\mu}(x)=\frac{1}{\Sigma(x)} \widetilde{e}_{k}^{\mu}.
\label{eq4}
\end{equation}
Here, the subscript $k = (1, 2, 3)$ stands for the spatial coordinates. In the light of rainbow gravity, Einstein's field equations are modified, based on the assumption that the geometry of space-time is related to the energy of the particle under test. Therefore, all amounts included in the field equations of this theory of gravity are energy-dependent. Contrary to the traditional Einstein field equations, a set of alternative field equations is being incorporated, of which the following is a typical one:
\begin{equation}
G_{\mu\nu}(x) \equiv R_{\mu\nu}(x) - \frac{1}{2}g_{\mu\nu}(x)R(x) = k(x)T_{\mu\nu}(x),
\label{eq10}
\end{equation}
where $k(x) = 8 \pi G(x)$. These modifications are of great interest, as they allow us to better elucidate the relationship between gravity, high-energy physics, and the fundamental structure of spacetime.

The Rastall-Rainbow gravity model \cite{Mota:2019zln} evolves by combining modifications of Rastall gravity and Rainbow gravity. It introduces a unified framework that captures both the non-minimal coupling between gravity and matter, along with the energy-dependent behavior of particles. This model furnishes an expanded description of gravity that goes beyond the classical formulation of GR. In this unified formalism, the field equations are given by,
\begin{equation}
R_{\mu}{}^{\nu}(x) - \frac{\lambda}{2}\delta_{\mu}{}^{\nu}(x)R(x) = k(x) T_{\mu}{}^{\nu}(x)  
 \label{eq7}.
\end{equation}
Here, the Rastall parameter $\lambda$ is supposed to be independent of the particle energy under test. In particular, within the framework of Rastall-Rainbow gravity, one can construct relativistic stars  \cite{Mota:2022zbq,Das:2022vxq,Li:2023fux} or
 exotic objects such  as wormholes \cite{Tangphati:2023nwz}. 
In the forthcoming sections, we shall concentrate on illustrating the simplest form of wormhole geometry and analyzing its geometric properties. Here, we assume $G(x) = 1$ for our entire calculation.

%%%%%%%%%%%%%%%%%%%%%%%%%%%%%%%%%%%%%%%%%%%

\subsection{The wormhole geometry and its corresponding field equations}

To construct a shape function satisfying embedding class one and characterizing the geometry of a spherically 
symmetric and static spacetime, we start with the metric ansatz given by
\begin{equation}\label{71}
ds^2=-e^{\Phi (r)} dt^2 +e^{\Lambda (r)} dr^2 +r^2 d\theta^2 +r^2\sin^2\theta d\phi^2.
\end{equation}
The geometry of spacetime is governed by the functions $\Phi(r)$ and $\Lambda(r)$, which depend on the radial coordinate $r$. These functions control the gravitational field and are ultimately involved in identifying the non-zero Riemann curvature components ($\ref{71}$) corresponding to the spacetime configuration under consideration.
\begin{eqnarray}\label{3.2.a}
 & \mathcal{R}_{{0101}}=\frac{1}{4}\,{{\rm e}^{\Phi}} \left( -\Phi^{{\prime}}\Lambda^{{\prime}}+{\Phi^{{\prime}}}^{2}+2
\,\Phi^{{\prime\prime}} \right), \\ \label{3.2.b}
 & \mathcal{R}_{{0202}}=\frac{1}{2}\,r\Phi^{{\prime}}{{\rm e}^{\Phi-\Lambda}},\\ \label{3.2.c}
 & \mathcal{R}_{{1202}}=0,\\ \label{3.2.d}
 & \mathcal{R}_{{1303}}=0,\\ \label{3.2.e}
 & \mathcal{R}_{{1313}}=\frac{1}{2}\,\Lambda^{{\prime}}r \sin^2 \theta,\\ \label{3.2.f}
 & \mathcal{R}_{{2323}}={r}^{2} {\sin^2 \theta} \left( 1-{{\rm e}^{-\Lambda}} \right).
\end{eqnarray}
The components of Riemann curvature listed above satisfy the well-known Karmarkar condition,
\begin{eqnarray}\label{75}
\mathcal{R}_{{0202}}\mathcal{R}_{{1313}}=\mathcal{R}_{{0101}}\mathcal{R}_{{2323}}+\mathcal{R}_{{1202}}\mathcal{R}_{{1303}}.\label{3.2}
\end{eqnarray}
In the space-time geometry environment, the embedding class-one concept relates to a very specific form in which the non-zero components of the Riemann curvature tensor can be embedded in a higher-dimensional flat space. This concept enables us to interpret curvature geometrically by means of an embedding function. Following that, we apply the Karmarkar condition by replacing the non-zero components of the Riemann curvature tensor. More precisely, in the case of a static spherically symmetric spacetime, the non-zero components are generally given by,
\begin{equation}\label{76}
\frac{\Phi' \Lambda'}{1-e^{\Lambda}}=\Phi' \Lambda'-2 \Phi''-\Phi'^2,
\end{equation}
where primes denote derivatives with respect to the radial coordinate $r$, and $e^{\Lambda}\neq1$. The above-mentioned differential equation (\ref{76}) gives the following solution,
\begin{equation}\label{78}
e^{\Lambda}=1+ A e^{\Phi} \Phi'^2.
\end{equation}
Here $A$ stands for integration constant. In exploring the properties of wormhole solutions, we start by considering a metric that is spherically symmetric, and time-independent (static). This metric substitutes the standard quantities of GR denoted by $\widetilde{e}_{i}$ for spherical symmetry in Eqn. (\ref{eq3}), leading to the following line element, 
\begin{equation}
    ds^{2}=-\mathcal{A}(r) dt^{2} + \mathcal{B}(r) dr^{2} + \frac{r^{2}}{\Sigma^{2}(x)}(d\theta^{2}+\sin{\theta}^{2}d\phi^{2}),
    \label{eq.24}
\end{equation}
where $\mathcal{A}(r)\equiv \frac{e^{2\Phi(r)}}{\Pi^{2}(x)}$ and $\mathcal{B}(r)\equiv \frac{1}{\Sigma^{2}(x) \left(1-\frac{b(r)}{r}\right)}$. In this context, $\Phi(r)$ and $b(r)$ serve as arbitrary functions of the radial coordinate and are called redshift and shape functions respectively. It should be noted that the metric potentials are influenced by the rainbow functions $\Pi(x)$ and $\Sigma(x)$. Nevertheless, it's worth mentioning that the coordinates $r$, $t$, $\theta$, and $\phi$ remain independent of the energy probe particles in gravity's rainbow.

A fundamental feature of wormhole physics concerns the occurrence of a throat connecting two asymptotically flat spacetimes, with a minimum surface radius denoted by $b(r_0) = r_0$. The radial coordinate covers the range between two patches, i.e. $[r_0, +\infty)$, where $r_0$ stands for the transition point between the patches. Moreover, the flaring-out condition plays a major role in wormhole's geometry and can be stated as follows \cite{Morris:1988cz},  
\begin{eqnarray}
\frac{b(r)-rb^{\prime}(r)}{b^2(r)}>0.
\end{eqnarray}
In more precise terms, the shape function, indicated by $b(r)$, should comply with the condition $b^{\prime}(r_0) < 1$ to maintain the well-defined nature of the throat. What's more, an auxiliary condition is also necessary for the region lying outside the throat, namely 
\begin{eqnarray}
1 - \frac{b(r)}{r} > 0.
\end{eqnarray}
 This condition guarantees that the space-time in the region surrounding the throat remains traversable. Importantly, the redshift function called $\Phi(r)$, must remain finite throughout spacetime, thereby avoiding the appearance of event horizons in the wormhole geometry. 

Our main concern is to develop a model that simulates a wormhole solution. In this regard, we generally employ an anisotropic fluid stress-energy tensor for describing the matter distribution, which is commonly described by,
\begin{equation}\label{eq12}
T_{\mu\nu}=(\rho+p_t)u_\mu u_\nu+ p_t g_{\mu\nu}-(p_{t}-p_{r}) \chi_{\mu}\chi_{\nu},
\end{equation}
where $u^\mu$ is the fluid 4-velocity denoted as $u_\mu u^\mu = -1$, $\chi_{\mu}$ is the unit radial vector satisfying the condition $\chi_{\mu} \chi^{\mu} = 1$ and $g_{\mu\nu}$ is the metric tensor. Furthermore, in this context, the energy density is represented by $\rho = \rho(r)$, where $r$ denotes the radial coordinate. Moreover, we have the radial pressure $p_r = p_r(r)$ and the tangential pressure $p_t = p_t(r)$, which depend solely on the radial coordinate, $r$.

Now, let us express Eq. (\ref{eq.4}) in its covariant form. On the left-hand side, we construct the geometrical Einstein tensor. On the right-hand side, we introduce an effective stress-energy tensor.
\begin{equation}
     R_{\mu\nu}-\frac{1}{2}g_{\mu\nu}R=8\pi G\tau_{\mu \nu},
     \label{eq13b}
 \end{equation}
  where
 \begin{equation}
     \tau_{\mu\nu}=T_{\mu\nu}-\frac{(1-\lambda)}{2(1-\lambda)}g_{\mu\nu}T.
      \label{eq13c}
 \end{equation}
Bearing in mind Eqs. (\ref{eq.5}) and (\ref{eq12}), the non-zero components of the field equations (\ref{eq13b}) can be reduced to 
  \begin{eqnarray}
8\pi  \Bar{\rho}&=& \frac{b^{\prime}}{r^{2}},  \label{eq14} \\
8\pi  \Bar{p}_{r} &=& 2\left(1-\frac{b}{r}\right)\frac{\Phi^{\prime}}{r}-\frac{b}{r^{3}} ,  \label{eq15} \\
8\pi  \Bar{p}_{t} &=& \left(1-\frac{b}{r}\right)\Big[\Phi^{\prime\prime}+\Phi^{\prime 2}-\frac{b^{\prime}r-b}{2r(r-b)}\Phi^{\prime}-\frac{b^{\prime}r-b}{2r^2(r-b)} +\frac{\Phi^{\prime}}{r}\Big]. \label{eq16}
\end{eqnarray}
Here, $\Bar{\rho}$, $\Bar{p}_{r}$, and $\Bar{p}_{t}$ stand for effective energy density, effective radial pressure, and effective tangential pressure respectively. One define,
\begin{align}
    \Bar{\rho} & = \frac{1}{\Sigma(x)^{2}}\left[\alpha_{1}\rho+\alpha_{2}p_{r}+2\alpha_{2}p_{t}\right],\label{eq17}\\
    \Bar{p}_{r} & = \frac{1}{\Sigma(x)^{2}}\left[\alpha_{2}\rho+\alpha_{1}p_{r}-2\alpha_{2}p_{t}\right], \label{eq18}\\
    \Bar{p}_{t} & = \frac{1}{\Sigma(x)^{2}}\left[\alpha_{2}\rho-\alpha_{2}p_{r}+\alpha_{3}p_{t}\right], \label{eq18b}
\end{align}
with
\begin{equation*}
    \alpha_{1}=\frac{1-3\lambda}{2(1-2\lambda)}, \qquad \alpha_{2}=\frac{1-\lambda}{2(1-2\lambda)}, \qquad
    \alpha_{3}=-\frac{\lambda}{1-2\lambda}.
\end{equation*}
When setting $\lambda=1$ and $\sigma=1$, the field equations (\ref{eq14})-(\ref{eq16}) can be modified to restore the standard GR definition for an anisotropic fluid sphere. In this modified formulation, there are five unknown quantities: $\Phi(r)$, $ b(r)$, $\rho(r)$, $p_r(r)$, and $p_t(r)$, which are governed by three independent differential equations. Nevertheless, this system of equations is undetermined, implying that additional information or suppositions are required to completely capture the wormhole configuration. In our peculiar approach, we have chosen to concentrate on establishing the functions $\Phi(r)$ and $b(r)$ as input variables. By doing so, we seek to get a somewhat more specific view of the wormhole. Lastly, it's worth mentioning the conservation equation of the stress-energy tensor $T^{\nu}_{\:\:\:\mu;\nu}=\Bar{\lambda}R_{\mu}$ takes the following form
\begin{equation}
 \Bar{p}_{r}' = -(\Bar{p}_{r}+\Bar{\rho})\Phi'+ \frac{2}{r}\left(\Bar{p}_{t}-\Bar{p}_{r}\right).
 \label{eq21}
\end{equation}

%%%%%%%%%%%%%%%%%%%%%%%%%%%%%%%%%%%%%%%%%%%%%%%%%%
 \section{Traversable wormhole geometry}\label{Sec3}
In order to construct a wormhole solution, we consider a simple form of redshift function $\Phi(r)$ that ensure the space-time is free from horizons and singularities as follows \cite{Kar:1995vm},
\begin{equation}\label{710}
\Phi (r)= -\frac{2 \xi}{r}.
\end{equation}
Here $\xi$ represents an arbitrary constant. By comparing equations ($\ref{71}$) and ($\ref{eq.24}$), one obtains the following result,
\begin{equation}\label{711}
\Lambda(r)= \ln \bigg[\frac{1}{\Sigma^{2}(x)}\frac{r}{r-b(r)}\bigg].
\end{equation}
The wormhole shape function, denoted $b(r)$, can be obtained by combining Eqs. ($\ref{78}$), ($\ref{710}$) and ($\ref{711}$), respectively. This enables us to derive the following expression for the wormhole shape function,
\begin{equation}\label{712}
b(r)= r - \frac{1}{\Sigma^{2}(x)}\frac{r^5}{r^4 + 4 \xi^2 A e^{\frac{-2 \xi}{r}}}.
\end{equation}
As stated by Morris and Thorne \cite{Morris:1988cz}, in order to achieve a traversable wormhole solution, the shape function $b(r)$ must satisfy the following crucial properties:\\
\begin{enumerate}
\item $ b'(r)<1~~\text{at}~~r=r_0$,

\item $b(r)-r=0~~\text{at}~~r=r_0$,

\item $ \frac{b(r)}{r} \rightarrow 0~~\text{at}~~r\rightarrow\infty$, 

\item $ \frac{b(r)-r b'(r)}{b'(r)}>0~~\text{at}~~r=r_0$.
\end{enumerate}
The parameter $r_0$ corresponds to the wormhole throat radius, and the radial coordinate $r$ satisfies the condition $r_0 \leq r \leq \infty$. To ensure the existence of a traversable wormhole solution, it is crucial that the shape function should satisfies these properties. These properties ensure that geodesics can pass through the throat without encountering singularities or obstructions. However, it's worthwhile to note that fulfilling these fundamental properties is necessary but not sufficient to establish the physical viability of a wormhole solution. Additional factors, such as the energy conditions, stability analysis and the behavior of matter fields, must also be taken into account to construct physically plausible wormholes.

\begin{figure*}
\centering \epsfig{file=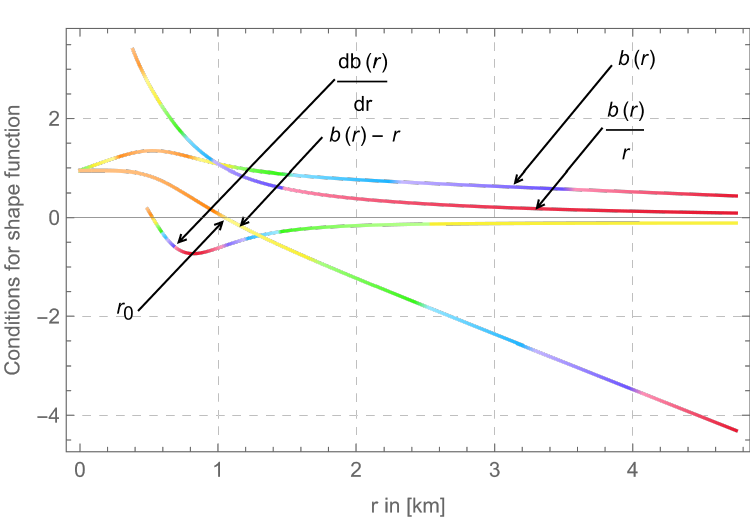, width=.35\linewidth,
height=2.1in}\centering \epsfig{file=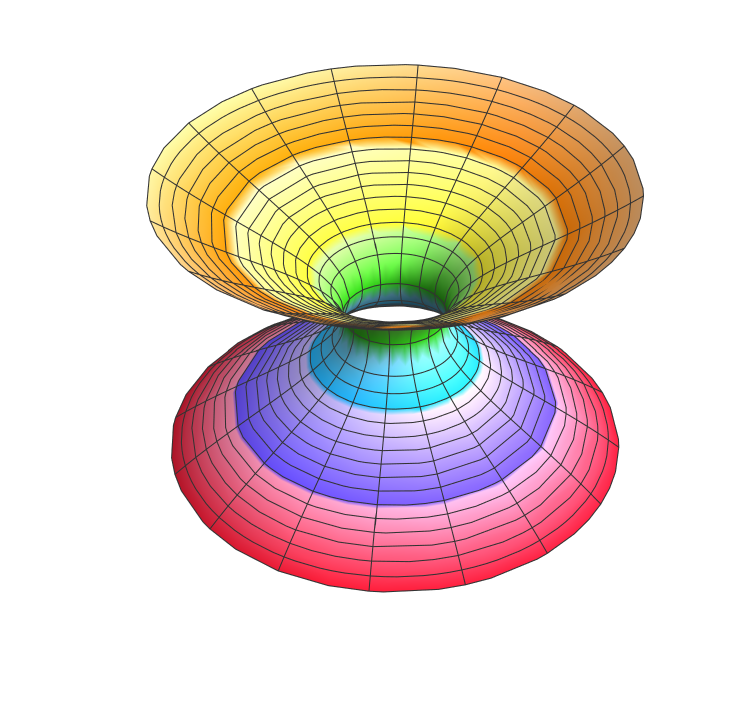, width=.35\linewidth,
height=2.1in}\centering \epsfig{file=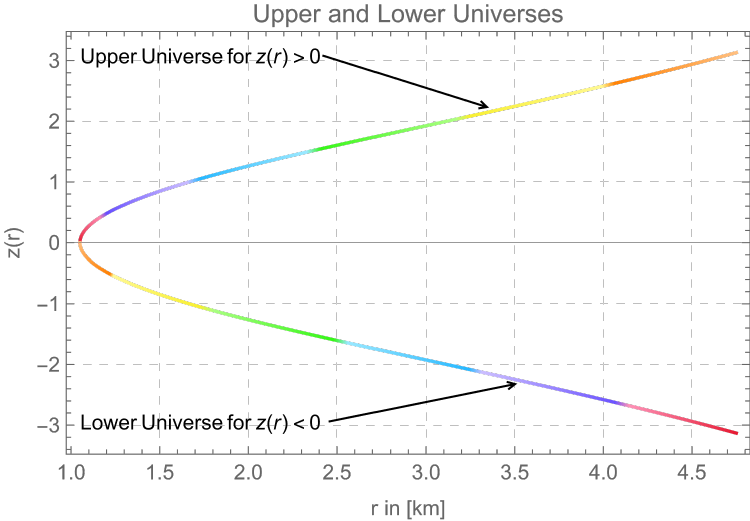, width=.35\linewidth,
height=2.1in}
\caption{\label{fig1} The plots presented here illustrate the properties of the shape function$-$(Left panel), the embedding diagram$-$(Middle panel) and the embedded surface $z(r)$$-$(Right panel) for our model. The wormhole is characterized by specific parameter values, namely $\Sigma(x) = 0.95$, $\gamma = 0.95$, $\xi =0.001$ and $r_0 = 1.05$.}
\end{figure*}

By substituting the throat condition $b(r_0) - r_0 = 0$ into Eq. ($\ref{712}$), we arrive at the trivial solution $r_0 = 0$. To address this problem, we introduce a free parameter $\gamma$ into Eq. ($\ref{712}$). Accordingly, Eq. ($\ref{712}$) has been modified to take the following form,
\begin{equation}
b(r)=r- \frac{1}{\Sigma^{2}(x)}\frac{r^5}{r^4 + 4 \xi^2 A e^{\frac{-2 \xi}{r}}}+\gamma.    
\end{equation}
Based on the information given by condition ($2$), we can infer the following,
\begin{equation}
A= \frac{r_{0}^{4}}{4 \gamma\Sigma^{2}(x) \xi^2 e^{\frac{-2\xi}{r_0}}}\left[r_0-\gamma\Sigma^{2}(x)\right].
 \end{equation}
Upon replacing the value of $A$ in Eq. ($\ref{712}$), we can derive an alternative formula for the shape function, $b(r)$. The resulting formula is as follows, 
\begin{equation}\label{713}
b(r)=r -\frac{\gamma  r^5}{r_0^4 \left(r_0-\gamma  \Sigma^{2}(x)\right) e^{-2 \xi  \left(\frac{1}{r}-\frac{1}{r_0}\right)}+\gamma \Sigma^{2}(x) r^4 }+\gamma,
\end{equation}
where $0<\gamma<r_0$. The prescribed range of $\gamma$ meets both conditions ($1$) and ($2$). By using condition ($3$) with Eq. ($\ref{713}$), we arrive at the following statement,
\begin{equation}\label{614}
\lim_{r \rightarrow \infty} \frac{b(r)}{r}=0.
\end{equation}
Consequently, by evaluating Eq. ($\ref{713}$), we get asymptotically flat traversable wormholes. To demonstrate each of the specified conditions, we have plotted a graph in Fig. $\ref{fig1}$, which clearly displays the fulfillment of all the necessary criteria by our generated wormhole. In addition, we have employed embedding diagrams to further visualize the wormhole geometry and extract valuable information. In the case of spherical symmetry, we are able to simplify the wormhole analysis by taking an equatorial slice with $\theta = \frac{\pi}{2}$. This choice capitalizes on the symmetry of the problem, in effect shrinking it to a two-dimensional slice. By holding the time coordinate at a constant value, $t = \text{const}$, we can investigate a snapshot of the wormhole geometry at a specific moment in time. This simplification allows us to concentrate on the spatial features of the wormhole without taking into account the dynamics related to temporal evolution. With $\theta = \frac{\pi}{2}$ and $t = \text{const}$, the wormhole metric can be expressed in terms of radial coordinates $r$ and azimuth angle $\phi$. Consequently, the resulting two-dimensional geometry can be represented by a line element of the following form, 
\begin{equation}\label{715}
ds^2=\frac{1}{\Sigma^{2}(x) \left(1-\frac{b(r)}{r}\right)} dr^2+r^2 d\phi^2.
\end{equation}
With cylindrical coordinates ($r, z, \phi$) in three-dimensional Euclidean space, we can visualize the line element ($\ref{715}$). Once visualized, this involves embedding the geometry of the wormhole in the selected coordinate system. The embedding enables us to portray the curved space-time of the wormhole as a surface in three-dimensional Euclidean space. The specific form of the embedding, depending on the line element ($\ref{715}$) and the peculiar wormhole model under consideration, can be stated by the following line element,
\begin{equation}\label{716}
ds^2= dr^2 + dz^2 +r^2 d\phi^2.
\end{equation}
In a three-dimensional Cartesian coordinate system, the embedded surface $z \equiv z(r)$ allows us to visualize the form of the wormhole with axial symmetry as follows,
\begin{equation}\label{717}
ds^2= \bigg[1+ \bigg(\frac{dz}{dr} \bigg)^2 \bigg]dr^2 +r^2 d\phi^2.
\end{equation}
By comparing Eqs. ($\ref{715}$) and ($\ref{717}$), we can easily determine that,
\begin{equation}\label{718}
\frac{dz}{dr}=\pm \bigg( \frac{1}{\Sigma^{2}(x) \left(1-\frac{b(r)}{r}\right)} -1  \bigg)^{\frac{1}{2}}.
\end{equation}
 
It's worth mentioning here that Eq. ($\ref{718}$) reveals an interesting characteristic: at the throat, the embedded surface becomes vertical, indicating that $\frac{dz}{dr} \rightarrow \infty$. This suggests a rapid change in the height coordinate as the radial coordinate approaches the throat, implying a significant curvature or narrowness in that region. Moreover, if $\frac{dz}{dr}$ tends to zero as $r$ approaches infinity, it implies that the height coordinate approaches a constant value away from the throat, resulting in an asymptotically flat behavior. This behavior is indicative of a spacetime that becomes increasingly similar to flat spacetime at large radial distances. The embedded diagram shown in Fig. $\ref{fig1}$ provides a visual representation of the wormhole geometry. Significantly, this diagram visualizes both the upper universe for $z>0$ and the lower universe for $z<0$. 

%------------------------------------------------------------
\section{Energy conditions}\label{Sec4}
 %------------------------------------------------------------
To ensure the physical consistency of solutions in the context of Einstein's equations, energy conditions (ECs) are imposed. These mathematical constraints serve as criteria for determining the validity of a spacetime geometry. The four primary ECs, namely the null energy condition (NEC), weak energy condition (WEC), dominant energy condition (DEC), and strong energy condition (SEC), play a crucial role in ensuring that the solutions satisfy certain physical requirements. By imposing these constraints on the energy-momentum tensor, we can assess the behavior of matter and energy within the framework of GR. According to classical GR, violations of the WEC are closely linked to the existence of exotic matter within wormhole structures. The violation of the WEC is characterized by the following condition
\begin{eqnarray}
    T_{\xi \varrho}u^{\xi} u^{\varrho}\geq 0,
\end{eqnarray}
where $T_{\xi \varrho}$ is the energy-momentum tensor, and $u^\xi$ represents any timelike vector. Hochberg \& Visser \cite{Hochberg:1998ha, Hochberg:1998ii}, building onto earlier work by Morris \& Thorne \cite{Morris:1988cz}, extensively investigated the violation of the NEC in the context of wormholes employing exotic matter. They showed that the throat region of a wormhole does indeed violate the NEC. The behavior of the $\Omega$ expansion, $w_{\xi \varrho}$ rotation, and $\sigma_{\xi \varrho}$ shear of the congruence associated with the $v^\xi$ vector field is governed by the Raychaudhuri equations, which can be expressed as follows,
\begin{eqnarray}
\frac{d \Omega}{d \tau} &=& - \frac{\Omega^2}{3}-R_{\xi \varrho} v^{\xi} v^{\varrho}-\sigma_{\xi \varrho} \sigma^{\xi \varrho}-w_{\xi \varrho} w^{\xi \varrho},\label{aa}\\
\frac{d \Omega}{d \tau} &=&- \frac{\Omega^2}{2}-R_{\xi \varrho} u^{\xi} u^{\varrho}-\sigma_{\xi \varrho} \sigma^{\xi \varrho}-w_{\xi \varrho} w^{\xi \varrho}.
\end{eqnarray}
In order for gravity to be attractive, the criterion is that $R_{\xi \varrho} v^{\xi} v^{\varrho}\geq0$ holds true for all hypersurface orthogonal congruences, where $w_{\xi \varrho}\equiv0$ and $\sigma^2=\sigma_{\xi \varrho} \sigma^{\xi \varrho}$. Mathematically, this condition can be expressed as, 
\begin{equation}
T_{\xi \varrho} v^{\xi} v^{\varrho}\geq0.
\end{equation}
The ECs, including the NEC, WEC, DEC, and SEC, are derived from the Einstein field equations. These conditions impose inequalities on the components of the energy-momentum tensor, which correspond to energy and momentum densities. Mathematically, they can be written as, 
 \begin{eqnarray}
NEC&:&~ (p_{k}+\rho)  \geq 0,\\
WEC&:&~\rho \geq0,~ (p_{k}+\rho ) \geq 0,\\
DEC&:&~ (\rho- |p_{k}| ) \geq 0,~\text{where $k =$ space index},\\
SEC&:&~ (\rho+ \overset{3}{\underset{k=1}{\Sigma}} p_{k})\geq 0,~ (p_{k}+\rho) \geq 0.
 \end{eqnarray}
A violation of the ECs is a strong indication of the presence of exotic matter. More specifically, wormhole configurations necessitate a violation of the NEC. It should be noted that the NEC is the weakest restrictive of the ECs. It should be noted that the NEC is the least restrictive of the energy conditions. It has been shown that the stability of static wormhole configurations relies on the existence of exotic matter. Importantly, however, no experimental evidence has been discovered to prove the existence of exotic matter, raising doubts about the validity and physical achievability of wormhole. 
%------------------------------------------------------------------

%----------------------------------------------------------
 \section{Exact solutions for the matter content of the wormhole in Rastall-Rainbow formalism}\label{Sec5}
%----------------------------------------------------------
\subsection{Analyzing the properties of shape function}

Now, we discuss the recently derived shape function, denoted by Eq. ($\ref{713}$). The zero tidal solution for the wormhole remains completely unaffected by the rainbow function $\Sigma(x)$, pointing to its robustness independently of the specific form of $\Sigma(x)$. It's also vital to ensure that the conditions $1 - \frac{b(r)}{r} > 0$ for $r > r_0$, $b'(r_0) < 1$, and $b(r) - b'(r) r > 0$ are satisfied, as previously noted. Upon examining Fig. \ref{fig1}, it becomes apparent that for $r > r_0$, the expression $b(r) - r$ takes on negative values, indicating that $b(r)/r < 1$. This implies that with $r \geq r_0$ and $b'(r) < 1$, the function $b(r) - r$ exhibits a decreasing trend with respect to $r$, satisfying the necessary flaring-out condition. Consequently, the derived shape function accurately characterizes the structure of a wormhole. Furthermore, Fig. \ref{fig1} illustrates the embedding diagram for wormhole, featuring both the upper universe (for $z > 0$) and the lower universe (for $z < 0$).  This is also verified that the spacetime is asymptotically flat, meaning that $\frac{dz}{dr}$ tends to zero as $r$ approaches infinity. 

\subsection{Analyzing the adiabatic sound velocity}
The stability and physical significance of the solutions can be efficiently evaluated by considering the adiabatic sound velocity, expressed as,
\begin{equation}
v_s^2 = \frac{\partial{<p>}}{\partial{\rho}},
\end{equation}
where $<p>$ represents the average pressure across the three spatial dimensions, defined as,
\begin{equation}
<p> = \frac{1}{3}(p_r + 2p_t).
\end{equation}
These fundamental properties hold true while respecting the following constraint, 
\begin{equation}
0 \leq v_s^2 < 1.
\end{equation}
By carefully analyzing Eqs. (\ref{rho}), (\ref{p_r}), and (\ref{p_t}), we can extract the following,
\begin{small}
\begin{eqnarray}\label{Sound}
v_s^2 &=& \frac{d<p>}{dr} \left( \frac{d\rho}{dr}\right)^{-1} = -\frac{1}{3} \Bigg[\mathcal{H}_1(r)\mathcal{G}_1^{-1}(r)+\mathcal{H}_2(r)\mathcal{G}_2^{-1}(r)\Bigg].
\end{eqnarray}
\end{small}
where,
\begin{small}
\begin{eqnarray*}
\mathcal{H}_1(r)&=&\lambda  (2 \lambda -1) \Sigma[x] ^2 \Bigg[\gamma ^2 r^8 \Sigma[x] ^2 \left(\gamma  \Sigma[x] ^2+2 r \left(\Sigma[x] ^2-1\right)\right) +2 \gamma  r^4 r_0^4 e^{2 \xi  \left(\frac{1}{r_0}-\frac{1}{r}\right)} \left(r_0-\gamma  \Sigma[x] ^2\right) \Big(\gamma  \Sigma[x] ^2+\xi \\&+&r \left(2 \Sigma[x] ^2-3\right)\Big)+r_0^8 (\gamma +2 r) e^{4 \xi  \left(\frac{1}{r_0}-\frac{1}{r}\right)} \left(r_0-\gamma  \Sigma[x] ^2\right)^2\Bigg],\\
\mathcal{H}_2(r)&=&(2 \lambda -1) \Sigma[x] ^2 \Big(\gamma ^2 r^8 \Sigma[x] ^2 \Bigg[\gamma  (3 \lambda -1) \Sigma[x] ^2+2 \lambda ^2 r \left(\Sigma[x] ^2-1\right)\Big)+2 \gamma  r^4 r_0^4 e^{2 \xi  \left(\frac{1}{r_0}-\frac{1}{r}\right)} \left(r_0-\gamma  \Sigma[x] ^2\right) \\&\times&\Big(\gamma  (3 \lambda -1) \Sigma[x] ^2+\left(2 \lambda ^2-3 \lambda +1\right) \xi + r \left(\lambda ^2 \left(2 \Sigma[x] ^2-5\right)+6 \lambda -2\right)\Big)+ r_0^8 \Big(\gamma  (3 \lambda -1)\\&+&2 \lambda ^2 r\Big) e^{4 \xi  \left(\frac{1}{r_0}-\frac{1}{r}\right)} \left(r_0-\gamma  \Sigma[x] ^2\right)^2\Bigg],
\end{eqnarray*}
\end{small}
\begin{small}
\begin{eqnarray*}
\mathcal{G}_1(r)&=&8 \pi  \left(\lambda ^2-3 \lambda +1\right) r^3 \Bigg[\gamma  r^4 \Sigma[x] ^2 +r_0^4 e^{2 \xi  \left(\frac{1}{r_0}-\frac{1}{r}\right)} \left(r_0-\gamma  \Sigma[x] ^2\right)\Bigg]^2,\\
\mathcal{G}_2(r)&=&16 \pi  \lambda  \left(\lambda ^2-3 \lambda +1\right) r^3 \Bigg[\gamma  r^4 \Sigma[x] ^2+r_0^4 e^{2 \xi  \left(\frac{1}{r_0}-\frac{1}{r}\right)} \left(r_0-\gamma  \Sigma[x] ^2\right)\Bigg]^2.
\end{eqnarray*}
\end{small}

\begin{figure*}
\centering \epsfig{file=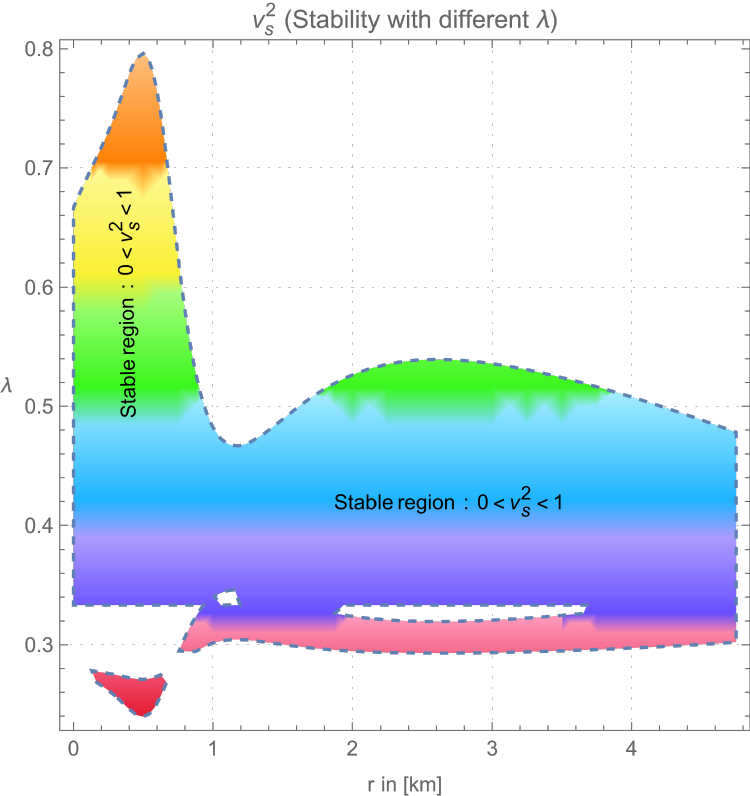, width=0.39\linewidth,
height=2.1in}~~~~~~~~\centering \epsfig{file=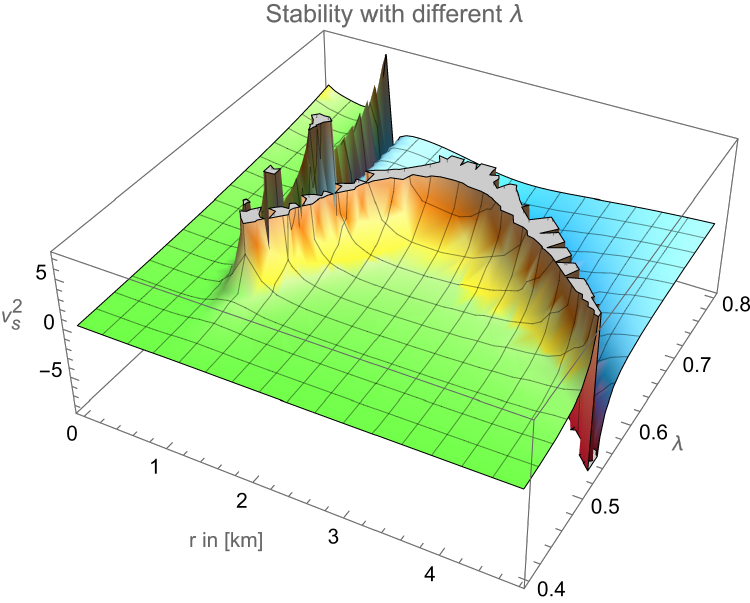, width=0.39\linewidth,
height=2.1in}
\caption{\label{fig2} The graph consists of two panels: the left panel displays a contour plot, while the right panel presents a three-dimensional plot. It illustrates the behavior of the parameter space ($\lambda$, $r$), and highlights the region where the adiabatic sound speed condition is satisfied. The wormhole solution is characterized by specific parameter values, namely $\Sigma(x) = 0.95$, $\gamma = 0.95$, $\xi =0.001$ and $r_0 = 1.05$.}
\end{figure*}

Thus, we see that sound velocity depends both on the parameters $\lambda$ and $\Sigma[x]$. Noted that the parameter $\lambda$ must be constrained within the regions where both the flaring-out and stability conditions are simultaneously satisfied. Fig. $\ref{fig2}$ illustrates the parameter space, highlighting the valid region for satisfying these conditions.
%-----------------------------------------------------------

\subsection{Generating the exact solutions}

Now, by substituting Eq. (\ref{713}) into the Eqs. (\ref{eq14})-(\ref{eq16}), the energy density and pressure components are readily derived,
\begin{small}
\begin{eqnarray}
    \rho(r)&=&-\mathcal{H}_3(r)\mathcal{G}_3^{-1}(r),\label{rho}\\
    p_r(r)&=&-\mathcal{H}_4(r)\mathcal{G}_4^{-1}(r),\label{p_r}\\
    p_t(r)&=&-\mathcal{H}_5(r)\mathcal{G}_5^{-1}(r)\label{p_t},
\end{eqnarray}
\end{small}
where,
\begin{small}
\begin{eqnarray*}
\mathcal{H}_3(r)&=& (1-2 \lambda ) \Sigma[x] ^2 \Bigg[(2 (1-\lambda ) \lambda ^2 \Big(\gamma ^2 r^8 \Sigma[x] ^2 \Big(\gamma  \Sigma[x] ^2+2 r \left(\Sigma[x] ^2-1\right)\Big)+2 \gamma  r^4 r_0^4 e^{2 \xi  \left(\frac{1}{r_0}-\frac{1}{r}\right)} \left(r_0-\gamma  \Sigma[x] ^2\right)\\&\times& \left(\gamma  \Sigma[x] ^2+\xi +r \left(2 \Sigma[x] ^2-3\right)\right)+r_0^8 (\gamma +2 r) e^{4 \xi  \left(\frac{1}{r_0}-\frac{1}{r}\right)} \left(r_0-\gamma  \Sigma[x] ^2\right)^2\Big)+(1-\lambda ) \Big(\gamma ^2 r^8 \Sigma[x] ^2 \Big(\gamma  (3 \lambda -1) \\&\times&\Sigma[x] ^2+2 \lambda ^2 r \left(\Sigma[x] ^2-1\right)\Big)+2 \gamma  r^4 r_0^4 e^{2 \xi  \left(\frac{1}{r_0}-\frac{1}{r}\right)} \left(r_0-\gamma  \Sigma[x] ^2\right) \Big(\gamma  (3 \lambda -1) \Sigma[x] ^2+\left(2 \lambda ^2-3 \lambda +1\right) \xi \\&+& r \left(\lambda ^2 \left(2 \Sigma[x] ^2-5\right)+6 \lambda -2\right)\Big)+r_0^8 \Big(\gamma  (3 \lambda -1)+2 \lambda ^2 r\Big) e^{4 \xi  \left(\frac{1}{r_0}-\frac{1}{r}\right)} \left(r_0-\gamma  \Sigma[x] ^2\right)^2\Big)+4 \left(\lambda ^2-3 \lambda +1\right) \\&\times&\lambda  r \Big(\gamma ^2 r^8 \Sigma[x] ^2+5 \gamma  r^4 r_0^4 e^{2 \xi  \left(\frac{1}{r_0}-\frac{1}{r}\right)} \left(r_0-\gamma  \Sigma[x] ^2\right)-\left(\gamma  r^4 \Sigma[x] ^2+r_0^4 e^{2 \xi  \left(\frac{1}{r_0}-\frac{1}{r}\right)} \left(r_0-\gamma  \Sigma[x] ^2\right)\right)^2\\&-&2 \gamma  \xi  r^3 r_0^4 e^{2 \xi  \left(\frac{1}{r_0}-\frac{1}{r}\right)} \left(r_0-\gamma  \Sigma[x] ^2\right)\Big)\Bigg],\\
\mathcal{G}_3(r)&=&16 \pi  (1-3 \lambda ) \lambda  \left(\lambda ^2-3 \lambda +1\right) r^3 \Bigg[\gamma  r^4 \Sigma[x] ^2+r_0^4 e^{2 \xi  \left(\frac{1}{r_0}-\frac{1}{r}\right)} \left(r_0-\gamma  \Sigma[x] ^2\right)\Bigg]^2,
\end{eqnarray*}
\end{small}
%-----------------------------------------------------------
\begin{figure*}
\centering \epsfig{file=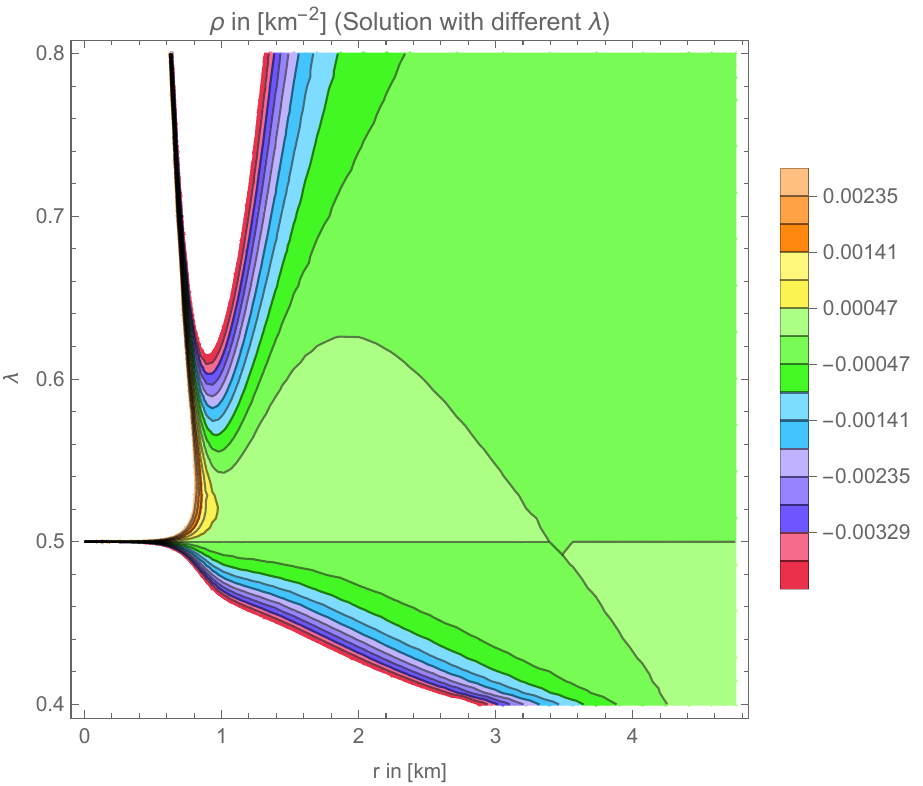, width=.35\linewidth,
height=2.1in}~~~~~~~~\centering \epsfig{file=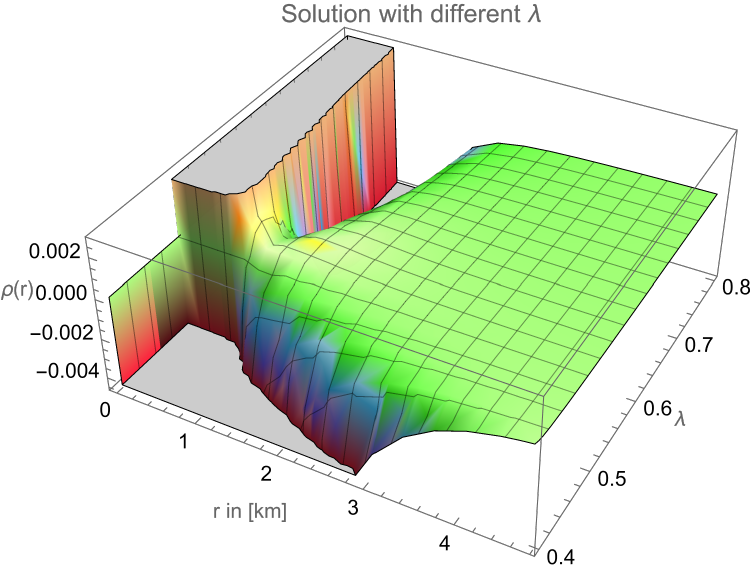, width=.35\linewidth,
height=2.1in}
\caption{\label{fig3} The graph consists of two panels: the left panel displays a contour plot, while the right panel presents a three-dimensional plot. It illustrates the behavior of the energy density ($\rho$) for our wormhole solution with different values of $\lambda$ in the range of $\lambda \in [0.4, 0.8]$. The wormhole is characterized by specific parameter values, namely $\Sigma(x) = 0.95$, $\gamma = 0.95$, $\xi =0.001$ and $r_0 = 1.05$.}
\end{figure*}
\begin{small}
\begin{eqnarray*}
\mathcal{H}_4(r)&=& (2 \lambda -1) \Sigma[x] ^2 \Bigg[\gamma ^2 r^8 \Sigma[x] ^2 \left(\gamma  (3 \lambda -1) \Sigma[x] ^2+2 \lambda ^2 r \left(\Sigma[x] ^2-1\right)\right)+2 \gamma  r^4 r_0^4 e^{2 \xi  \left(\frac{1}{r_0}-\frac{1}{r}\right)} \left(r_0-\gamma  \Sigma[x] ^2\right) \Big(\gamma  (3 \lambda -1) \Sigma[x] ^2\\&+&\left(2 \lambda ^2-3 \lambda +1\right) \xi +r \left(\lambda ^2 \left(2 \Sigma[x] ^2-5\right)+6 \lambda -2\right)\Big)+r_0^8 \left(\gamma  (3 \lambda -1)+2 \lambda ^2 r\right) e^{4 \xi  \left(\frac{1}{r_0}-\frac{1}{r}\right)} \left(r_0-\gamma  \Sigma[x] ^2\right)^2\Bigg],\\
\mathcal{G}_4(r)&=& 16 \pi  \lambda  \left(\lambda ^2-3 \lambda +1\right) r^3 \Bigg[\gamma  r^4 \Sigma[x] ^2+r_0^4 e^{2 \xi  \left(\frac{1}{r_0}-\frac{1}{r}\right)} \left(r_0-\gamma  \Sigma[x] ^2\right)\Bigg]^2,
\end{eqnarray*}
\end{small}
\begin{small}
\begin{eqnarray*}
\mathcal{H}_5(r)&=&\lambda  (2 \lambda -1) \Sigma[x] ^2 \Bigg[\gamma ^2 r^8 \Sigma[x] ^2 \left(\gamma  \Sigma[x] ^2+2 r \left(\Sigma[x] ^2-1\right)\right)+2 \gamma  r^4 r_0^4 e^{2 \xi  \left(\frac{1}{r_0}-\frac{1}{r}\right)} \left(r_0-\gamma  \Sigma[x] ^2\right) \Big(\gamma  \Sigma[x] ^2+\xi \\&+&r \left(2 \Sigma[x] ^2-3\right)\Big)+r_0^8 (\gamma +2 r) e^{4 \xi  \left(\frac{1}{r_0}-\frac{1}{r}\right)} \left(r_0-\gamma  \Sigma[x] ^2\right)^2\Bigg],\\
\mathcal{G}_5(r)&=& 16 \pi  \left(\lambda ^2-3 \lambda +1\right) r^3 \Bigg[\gamma  r^4 \Sigma[x] ^2+r_0^4 e^{2 \xi  \left(\frac{1}{r_0}-\frac{1}{r}\right)} \left(r_0-\gamma  \Sigma[x] ^2\right)\Bigg]^2.
\end{eqnarray*}
\end{small}

The correlation between energy density ($\rho(r)$), two pressure components ($p_r(r)$ and $p_t(r)$) and the Rainbow function $\Sigma(x)$ becomes obvious, highlighting the interdependence of these quantities. In Fig. \ref{fig3}, we depict the graphical behaviour of the energy density $\rho$ through specific parameter values: $\Sigma(x) = 0.95$, $\gamma = 0.95$, $\xi =0.001$ and $r_0 = 1.05$, respectively. This graph aims to display the validity of the inequality $\rho \geq 0$ for the wormhole solution over a range of values of $\lambda$, in particular in the region of $\lambda \in [0.4, 0.8]$. However, it's interesting to note that there's an exception in the region $\lambda \in [0.4, 0.8]$ where $\rho$ falls within the range $[-0.00414, 0]$, which renders it invalid for the admissible range of $\lambda$.
%--------------------------------------------------------------
\begin{figure*}
\centering \epsfig{file=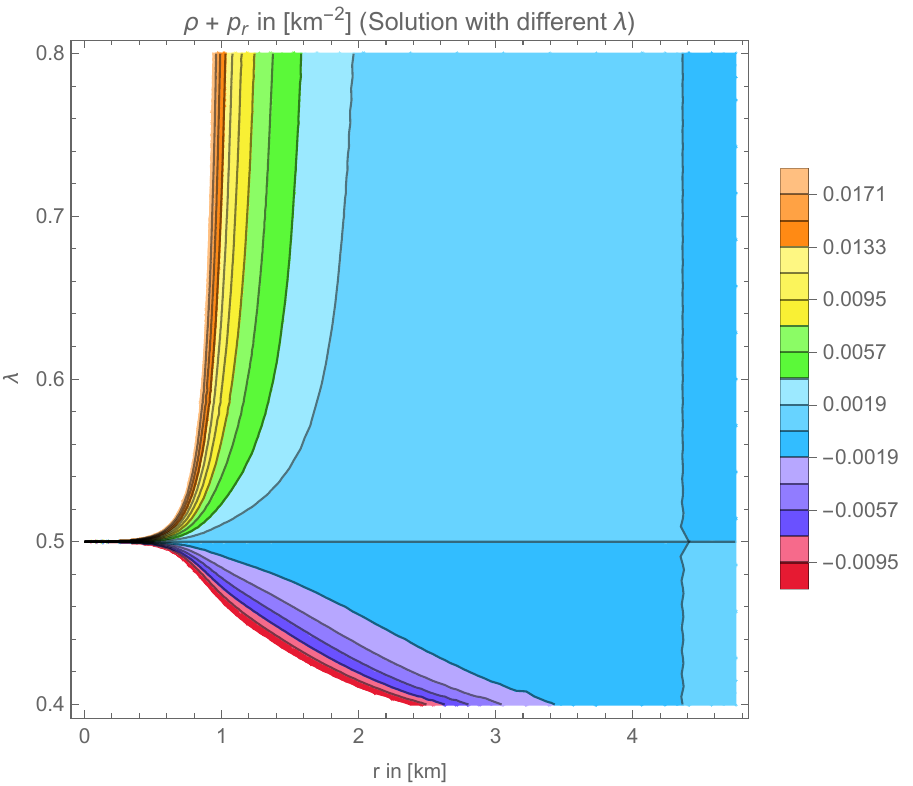, width=.35\linewidth,
height=2.1in}~~~~~~~~\centering \epsfig{file=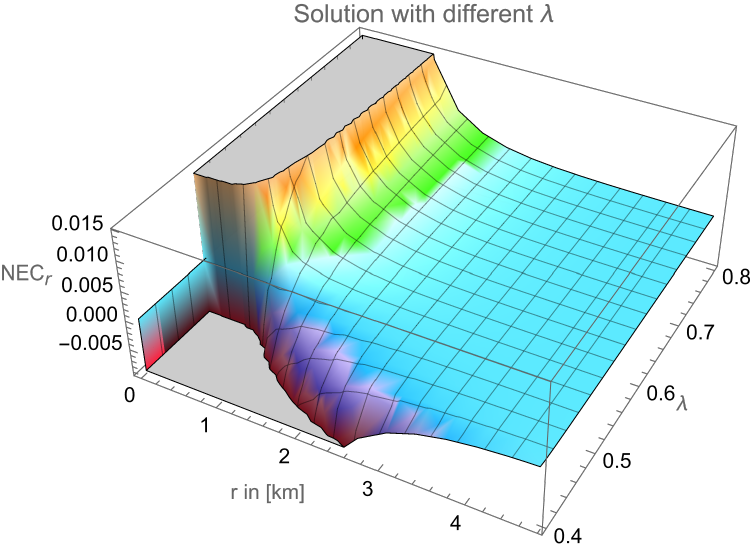, width=.35\linewidth,
height=2.1in}
\caption{\label{fig4} The graph consists of two panels: the left panel displays a contour plot, while the right panel presents a three-dimensional plot. It illustrates the behavior of the NEC (WEC) i.e., ($\rho+p_r$) for the wormhole solution with different values of $\lambda$ in the range of $\lambda \in [0.4, 0.8]$. The wormhole is characterized by specific parameter values, namely $\Sigma(x) = 0.95$, $\gamma = 0.95$, $\xi =0.001$ and $r_0 = 1.05$, respectively.}
\end{figure*}

This is more precisely stated by saying that what kind of matter contains the Rastall-Rainbow wormhole solution, and thus we search for different ECs in the forthcoming section. This helps us to assess ECs and draw a meaningful conclusion. The derived formulas for NEC, WEC, DEC and SEC are based on the Eqs. (\ref{rho})-(\ref{p_t}).
\begin{itemize}
    \item Searching for NEC and WEC along the radial and tangential directions which are indicated by
\begin{eqnarray}\label{e27}
\rho+p_r &=& \mathcal{H}_6(r)\mathcal{G}_6^{-1}(r),\\
\rho+p_t &=& \mathcal{H}_7(r)\mathcal{G}_7^{-1}(r)\label{e28}.
\end{eqnarray}
where,
\begin{small}
\begin{eqnarray*}
\mathcal{H}_6(r)&=&(2 \lambda -1) \left(\lambda ^2-4 \lambda +1\right) \Sigma[x] ^2 \Bigg[\gamma ^2 r^8 \Sigma[x] ^2 \left(\gamma  \Sigma[x] ^2+2 r \left(\Sigma[x] ^2-1\right)\right)+2 \gamma  r^4 r_0^4 e^{2 \xi  \left(\frac{1}{r_0}-\frac{1}{r}\right)} \left(r_0-\gamma  \Sigma[x] ^2\right) \\&\times&\left(\gamma  \Sigma[x] ^2+\xi +r \left(2 \Sigma[x] ^2-3\right)\right)+r_0^8 (\gamma +2 r) e^{4 \xi  \left(\frac{1}{r_0}-\frac{1}{r}\right)} \left(r_0-\gamma  \Sigma[x] ^2\right)^2\Bigg],\\
\mathcal{G}_6(r)&=&8 \pi  (3 \lambda -1) \left(\lambda ^2-3 \lambda +1\right) r^3 \Bigg[\gamma  r^4 \Sigma[x] ^2+r_0^4 e^{2 \xi  \left(\frac{1}{r_0}-\frac{1}{r}\right)} \left(r_0-\gamma  \Sigma[x] ^2\right)\Bigg]^2,\\
\mathcal{H}_7(r)&=&\Sigma[x] ^2 \Bigg[-\lambda ^2 (2 \lambda -1) \Big(\gamma ^2 r^8 \Sigma[x] ^2 \left(\gamma  \Sigma[x] ^2+2 r \left(\Sigma[x] ^2-1\right)\right)+2 \gamma  r^4 r_0^4 e^{2 \xi  \left(\frac{1}{r_0}-\frac{1}{r}\right)} \left(r_0-\gamma  \Sigma[x] ^2\right) \Big(\gamma  \Sigma[x] ^2+\xi \\&+& r \left(2 \Sigma[x] ^2-3\right)\Big)+r_0^8 (\gamma +2 r) e^{4 \xi  \left(\frac{1}{r_0}-\frac{1}{r}\right)} \left(r_0-\gamma  \Sigma[x] ^2\right)^2\Big)-\frac{1}{1-3 \lambda }\Bigg[(1-2 \lambda ) \Big(2 (1-\lambda ) \lambda ^2 \Big(\gamma ^2 r^8 \Sigma[x] ^2 \Big(\gamma  \Sigma[x] ^2 \\&+&2 r \left(\Sigma[x] ^2-1\right)\Big)+2 \gamma  r^4 r_0^4 e^{2 \xi  \left(\frac{1}{r_0}-\frac{1}{r}\right)} \left(r_0-\gamma  \Sigma[x] ^2\right) \left(\gamma  \Sigma[x] ^2+\xi +r \left(2 \Sigma[x] ^2-3\right)\right)+r_0^8 (\gamma +2 r) e^{4 \xi  \left(\frac{1}{r_0}-\frac{1}{r}\right)} \\&\times&\left(r_0-\gamma  \Sigma[x] ^2\right)^2\Big)+(1-\lambda ) \Big(\gamma ^2 r^8 \Sigma[x] ^2 \left(\gamma  (3 \lambda -1) \Sigma[x] ^2+2 \lambda ^2 r \left(\Sigma[x] ^2-1\right)\right)+2 \gamma  r^4 r_0^4 e^{2 \xi  \left(\frac{1}{r_0}-\frac{1}{r}\right)} \\&\times&\left(r_0-\gamma  \Sigma[x] ^2\right) \left(\gamma  (3 \lambda -1) \Sigma[x] ^2+\left(2 \lambda ^2-3 \lambda +1\right) \xi +r \left(\lambda ^2 \left(2 \Sigma[x] ^2-5\right)+6 \lambda -2\right)\right)+r_0^8 \Big(\gamma  (3 \lambda -1)\\&+&2 \lambda ^2 r\Big) e^{4 \xi  \left(\frac{1}{r_0}-\frac{1}{r}\right)} \left(r_0-\gamma  \Sigma[x] ^2\right)^2\Big)+4 \left(\lambda ^2-3 \lambda +1\right) \lambda  r \Big(\gamma ^2 r^8 \Sigma[x] ^2+5 \gamma  r^4 r_0^4 e^{2 \xi  \left(\frac{1}{r_0}-\frac{1}{r}\right)} \left(r_0-\gamma  \Sigma[x] ^2\right)\\&-&\left(\gamma  r^4 \Sigma[x] ^2+r_0^4 e^{2 \xi  \left(\frac{1}{r_0}-\frac{1}{r}\right)} \left(r_0-\gamma  \Sigma[x] ^2\right)\right)^2-2 \gamma  \xi  r^3 r_0^4 e^{2 \xi  \left(\frac{1}{r_0}-\frac{1}{r}\right)} \left(r_0-\gamma  \Sigma[x] ^2\right)\Big)\Big)\Bigg]\Bigg],\\
\mathcal{G}_7(r)&=& 16 \pi  \lambda  \left(\lambda ^2-3 \lambda +1\right) r^3 \left(\gamma  r^4 \Sigma[x] ^2+r_0^4 e^{2 \xi  \left(\frac{1}{r_0}-\frac{1}{r}\right)} \left(r_0-\gamma  \Sigma[x] ^2\right)\right)^2.
\end{eqnarray*}
\end{small}
    \item Searching for DEC and SEC along the radial and tangential directions which are indicated by
    \begin{eqnarray}
\rho-p_r &=&\mathcal{H}_8(r)\mathcal{G}_8^{-1}(r),\label{e29}\\
\rho-p_t &=& \mathcal{H}_9(r)\mathcal{G}_9^{-1}(r),\label{e30}
\\\label{e31}
\rho+p_r+2p_t &=& -\mathcal{H}_{10}(r)\mathcal{G}_{10}^{-1}(r),
\end{eqnarray}
where,
%------------------------
\begin{figure*}
\centering \epsfig{file=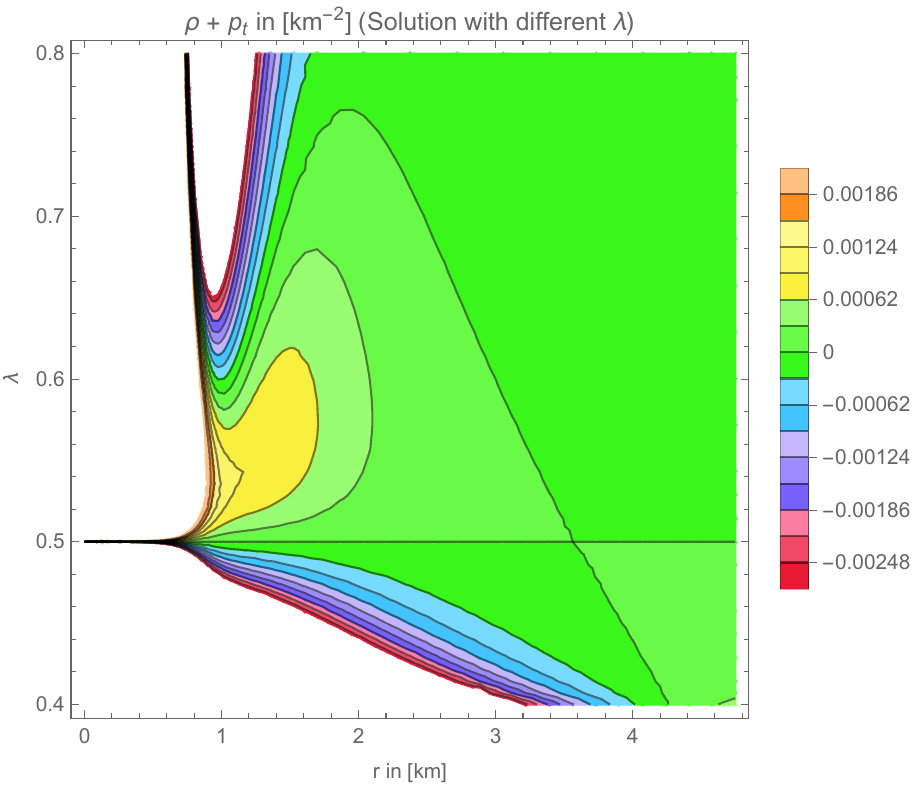, width=.35\linewidth,
height=2.1in}~~~~~~~~\centering \epsfig{file=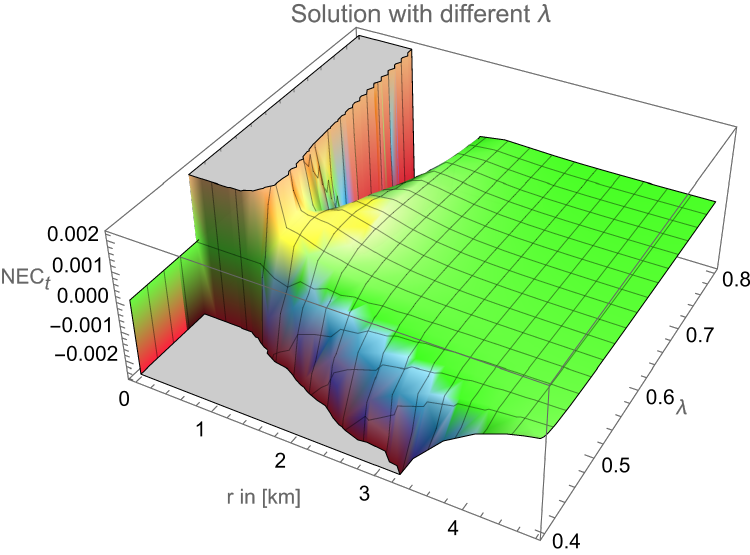, width=.35\linewidth,
height=2.1in}
\caption{\label{fig5} The graph consists of two panels: the left panel displays a contour plot, while the right panel presents a three-dimensional plot. It illustrates the behavior of the NEC (WEC) i.e., ($\rho+p_t$) for the wormhole solution with different values of $\lambda$ in the range of $\lambda \in [0.4, 0.8]$. The wormhole is characterized by specific parameter values, namely $\Sigma(x) = 0.95$, $\gamma = 0.95$, $\xi =0.001$ and $r_0 = 1.05$, respectively.}
\end{figure*}
%-----------------------------------------------
\begin{figure*}
\centering \epsfig{file=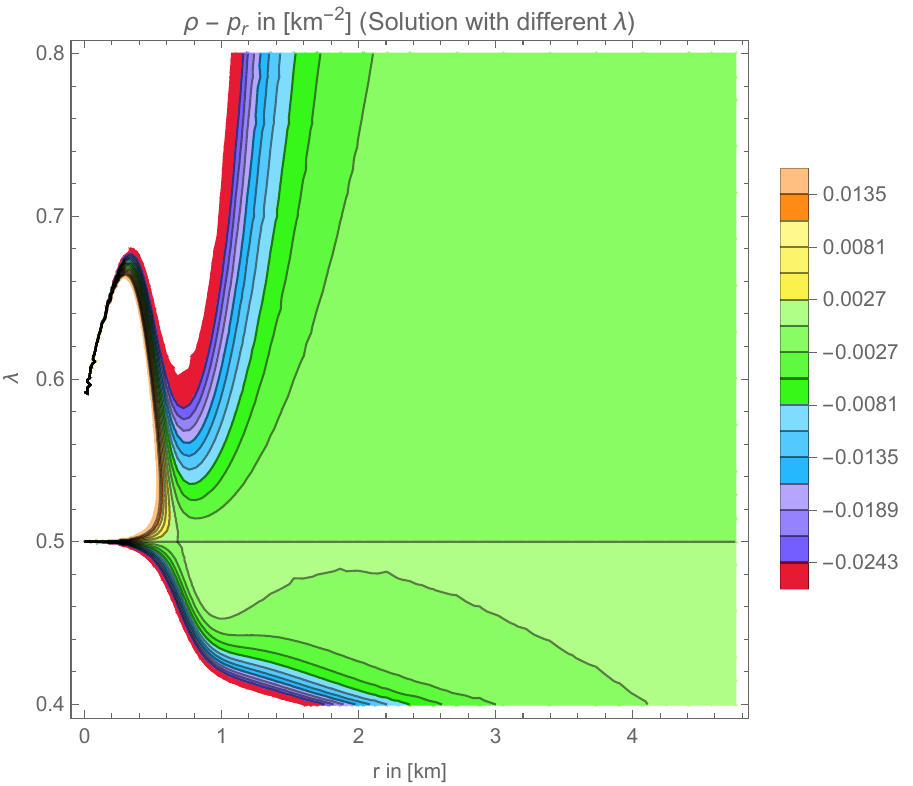, width=.35\linewidth,
height=2.1in}~~~~~~~~\centering \epsfig{file=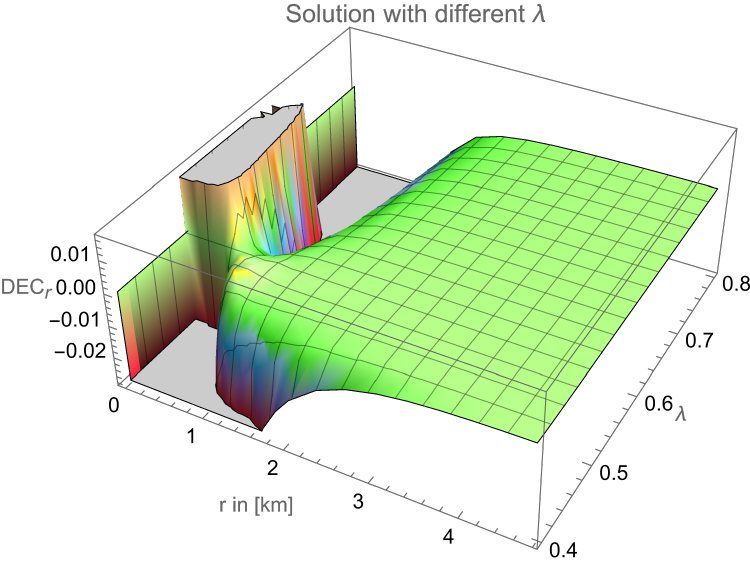, width=.35\linewidth,
height=2.1in}
\caption{\label{fig6} The graph consists of two panels: the left panel displays a contour plot, while the right panel presents a three-dimensional plot. It illustrates the behavior of the DEC i.e., ($\rho-p_r$) for our wormhole solution with different values of $\lambda$ in the range of $\lambda \in [0.4, 0.8]$. The wormhole was characterized by specific parameter values, namely $\Sigma(x) = 0.95$, $\gamma = 0.95$, $\xi =0.001$ and $r_0 = 1.05$, respectively.}
\end{figure*}
%----------------------------------------------------
\begin{small}
\begin{eqnarray*}
\mathcal{H}_8(r)&=&\Sigma[x] ^2 \Bigg[(2 \lambda -1) \Big(\gamma ^2 r^8 \Sigma[x] ^2 \Big(\gamma  (3 \lambda -1) \Sigma[x] ^2+2 \lambda ^2 r \left(\Sigma[x] ^2-1\right)\Big)+2 \gamma  r^4 r_0^4 e^{2 \xi  \left(\frac{1}{r_0}-\frac{1}{r}\right)} \left(r_0-\gamma  \Sigma[x] ^2\right) \\&\times& \left(\gamma  (3 \lambda -1) \Sigma[x] ^2+\left(2 \lambda ^2-3 \lambda +1\right) \xi +r \left(\lambda ^2 \left(2 \Sigma[x] ^2-5\right)+6 \lambda -2\right)\right)+r_0^8 \left(\gamma  (3 \lambda -1)+2 \lambda ^2 r\right) \\&\times& e^{4 \xi   \left(\frac{1}{r_0}-\frac{1}{r}\right)} \left(r_0-\gamma  \Sigma[x] ^2\right)^2\Big)-(1-3 \lambda)\Bigg[(1-2 \lambda ) \Big(2 (1-\lambda ) \lambda ^2 \Big(\gamma ^2 r^8 \Sigma[x] ^2 \left(\gamma  \Sigma[x] ^2+2 r \left(\Sigma[x] ^2-1\right)\right) \\&+&2 \gamma  r^4 r_0^4 e^{2 \xi  \left(\frac{1}{r_0}-\frac{1}{r}\right)} \left(r_0-\gamma  \Sigma[x] ^2\right) \left(\gamma  \Sigma[x] ^2+\xi +r \left(2 \Sigma[x] ^2-3\right)\right)+r_0^8 (\gamma +2 r) e^{4 \xi  \left(\frac{1}{r_0}-\frac{1}{r}\right)} \left(r_0-\gamma  \Sigma[x] ^2\right)^2\Big)\\&+&(1-\lambda ) \Big(\gamma ^2 r^8 \Sigma[x] ^2 \left(\gamma  (3 \lambda -1) \Sigma[x] ^2+2 \lambda ^2 r \left(\Sigma[x] ^2-1\right)\right)+2 \gamma  r^4 r_0^4 e^{2 \xi  \left(\frac{1}{r_0}-\frac{1}{r}\right)} \left(r_0-\gamma  \Sigma[x] ^2\right) \Big(\gamma  (3 \lambda -1) \Sigma[x] ^2\\&+&\left(2 \lambda ^2-3 \lambda +1\right) \xi +r \left(\lambda ^2 \left(2 \Sigma[x] ^2-5\right)+6 \lambda -2\right)\Big)+r_0^8 \left(\gamma  (3 \lambda -1)+2 \lambda ^2 r\right) e^{4 \xi  \left(\frac{1}{r_0}-\frac{1}{r}\right)} \left(r_0-\gamma  \Sigma[x] ^2\right)^2\Big)\\&+&4 \left(\lambda ^2-3 \lambda +1\right) \lambda  r \Big(\gamma ^2 r^8 \Sigma[x] ^2+5 \gamma  r^4 r_0^4 e^{2 \xi  \left(\frac{1}{r_0}-\frac{1}{r}\right)} \left(r_0-\gamma  \Sigma[x] ^2\right)-\Big(\gamma  r^4 \Sigma[x] ^2+r_0^4 e^{2 \xi  \left(\frac{1}{r_0}-\frac{1}{r}\right)} \\&\times&\left(r_0-\gamma  \Sigma[x] ^2\right)\Big)^2-2 \gamma  \xi  r^3 r_0^4 e^{2 \xi  \left(\frac{1}{r_0}-\frac{1}{r}\right)} \left(r_0-\gamma  \Sigma[x] ^2\right)\Big)\Big)\Bigg]\Bigg],\\
\mathcal{G}_8(r)&=& 16 \pi  \lambda  \left(\lambda ^2-3 \lambda +1\right) r^3 \Bigg[\gamma  r^4 \Sigma[x] ^2+r_0^4 e^{2 \xi  \left(\frac{1}{r_0}-\frac{1}{r}\right)} \left(r_0-\gamma  \Sigma[x] ^2\right)\Bigg]^2,
\end{eqnarray*}
\end{small}
\begin{small}
\begin{eqnarray*}
\mathcal{H}_9(r)&=&\Sigma[x] ^2 \Bigg[\lambda ^2 (2 \lambda -1) \Big(\gamma ^2 r^8 \Sigma[x] ^2 \left(\gamma  \Sigma[x] ^2+2 r \left(\Sigma[x] ^2-1\right)\right)+2 \gamma  r^4 r_0^4 e^{2 \xi  \left(\frac{1}{r_0}-\frac{1}{r}\right)} \left(r_0-\gamma  \Sigma[x] ^2\right) \Big(\gamma  \Sigma[x] ^2+\xi \\&+& r \left(2 \Sigma[x] ^2-3\right)\Big)+r_0^8 (\gamma +2 r) e^{4 \xi  \left(\frac{1}{r_0}-\frac{1}{r}\right)} \left(r_0-\gamma  \Sigma[x] ^2\right)^2\Big)-(1-3 \lambda)\Bigg[(1-2 \lambda ) \Big(2 (1-\lambda ) \lambda ^2 \Big(\gamma ^2 r^8 \Sigma[x] ^2\\&\times& \left(\gamma  \Sigma[x] ^2+2 r \left(\Sigma[x] ^2-1\right)\right)+2 \gamma  r^4 r_0^4 e^{2 \xi  \left(\frac{1}{r_0}-\frac{1}{r}\right)} \left(r_0-\gamma  \Sigma[x] ^2\right) \left(\gamma  \Sigma[x] ^2+\xi +r \left(2 \Sigma[x] ^2-3\right)\right)+r_0^8 (\gamma +2 r) \\&\times& e^{4 \xi  \left(\frac{1}{r_0}-\frac{1}{r}\right)} \left(r_0-\gamma  \Sigma[x] ^2\right)^2\Big)+(1-\lambda ) \Big(\gamma ^2 r^8 \Sigma[x] ^2 \left(\gamma  (3 \lambda -1) \Sigma[x] ^2+2 \lambda ^2 r \left(\Sigma[x] ^2-1\right)\right)+2 \gamma  r^4 r_0^4 e^{2 \xi  \left(\frac{1}{r_0}-\frac{1}{r}\right)} \\&\times& \left(r_0-\gamma  \Sigma[x] ^2\right) \left(\gamma  (3 \lambda -1) \Sigma[x] ^2+\left(2 \lambda ^2-3 \lambda +1\right) \xi +r \left(\lambda ^2 \left(2 \Sigma[x] ^2-5\right)+6 \lambda -2\right)\right)+r_0^8 \left(\gamma  (3 \lambda -1)+2 \lambda ^2 r\right)\\&\times& e^{4 \xi  \left(\frac{1}{r_0}-\frac{1}{r}\right)} \left(r_0-\gamma  \Sigma[x] ^2\right)^2\Big)+4 \left(\lambda ^2-3 \lambda +1\right) \lambda  r \Big(\gamma ^2 r^8 \Sigma[x] ^2+5 \gamma  r^4 r_0^4 e^{2 \xi  \left(\frac{1}{r_0}-\frac{1}{r}\right)} \left(r_0-\gamma  \Sigma[x] ^2\right)\\&-&\Big(\gamma  r^4 \Sigma[x] ^2+r_0^4 e^{2 \xi  \left(\frac{1}{r_0}-\frac{1}{r}\right)} \left(r_0-\gamma  \Sigma[x] ^2\right)\Big)^2-2 \gamma  \xi  r^3 r_0^4 e^{2 \xi  \left(\frac{1}{r_0}-\frac{1}{r}\right)} \left(r_0-\gamma  \Sigma[x] ^2\right)\Big)\Big)\Bigg]\Bigg],\\
\mathcal{G}_9(r)&=& 16 \pi  \lambda  \left(\lambda ^2-3 \lambda +1\right) r^3 \Bigg[\gamma  r^4 \Sigma[x] ^2+r_0^4 e^{2 \xi  \left(\frac{1}{r_0}-\frac{1}{r}\right)} \left(r_0-\gamma  \Sigma[x] ^2\right)\Bigg]^2,\\
\mathcal{H}_{10}(r)&=&(2 \lambda -1) \left(2 \lambda ^2+3 \lambda -1\right) \Sigma[x] ^2 \Bigg[\gamma ^2 r^8 \Sigma[x] ^2 \left(\gamma  \Sigma[x] ^2+2 r \left(\Sigma[x] ^2-1\right)\right)+2 \gamma  r^4 r_0^4 e^{2 \xi  \left(\frac{1}{r_0}-\frac{1}{r}\right)} \left(r_0-\gamma  \Sigma[x] ^2\right)\\&\times& \left(\gamma  \Sigma[x] ^2+\xi +r \left(2 \Sigma[x] ^2-3\right)\right)+r_0^8 (\gamma +2 r) e^{4 \xi  \left(\frac{1}{r_0}-\frac{1}{r}\right)} \left(r_0-\gamma  \Sigma[x] ^2\right)^2\Bigg],\\
\mathcal{G}_{10}(r)&=& 8 \pi  (3 \lambda -1) \left(\lambda ^2-3 \lambda +1\right) r^3 \Bigg[\gamma  r^4 \Sigma[x] ^2+r_0^4 e^{2 \xi  \left(\frac{1}{r_0}-\frac{1}{r}\right)} \left(r_0-\gamma  \Sigma[x] ^2\right)\Bigg]^2 .
\end{eqnarray*}
\end{small}
\end{itemize}

\begin{figure*}
\centering \epsfig{file=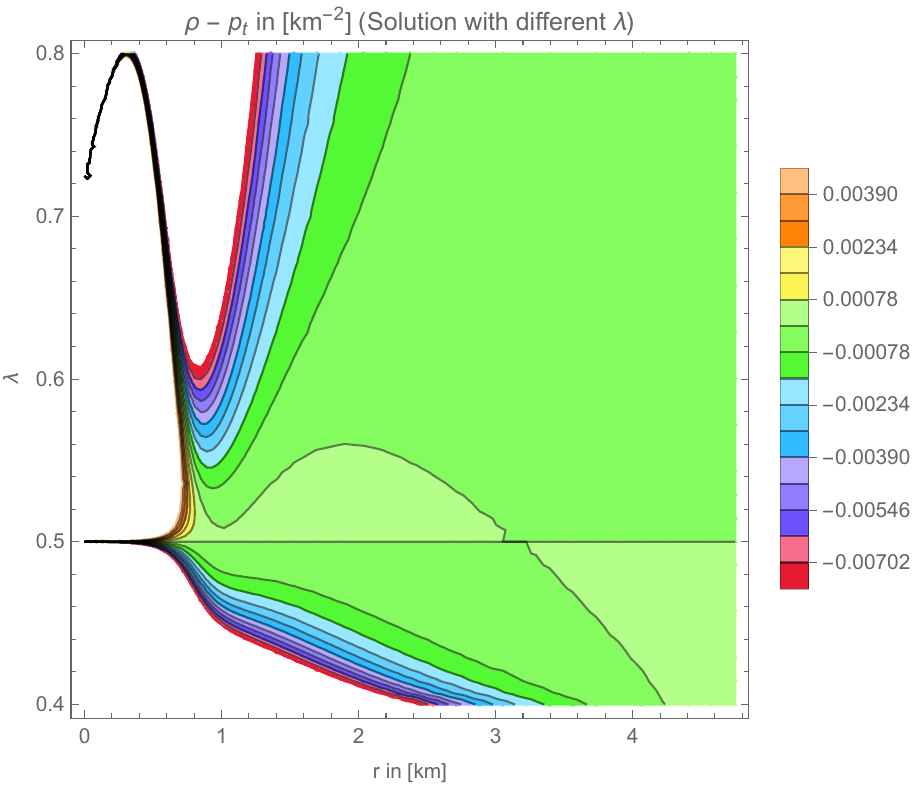, width=.35\linewidth,
height=2.1in}~~~~~~~~\centering \epsfig{file=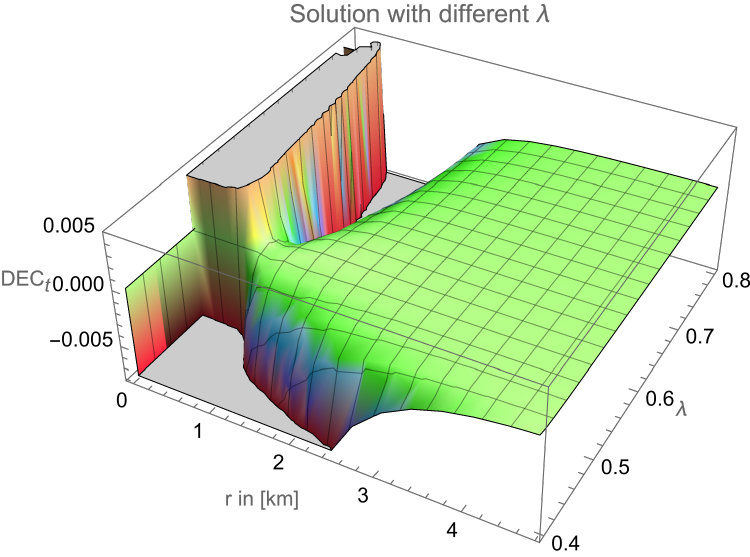, width=.35\linewidth,
height=2.1in}
\caption{\label{fig7} The graph consists of two panels: the left panel displays a contour plot, while the right panel presents a three-dimensional plot. It illustrates the behavior of the DEC i.e.,
($\rho-p_t$)  for our  wormhole solution with different values of $\lambda$ in the range of $\lambda \in [0.4, 0.8]$. The wormhole was characterized by specific parameter values, namely $\Sigma(x) = 0.95$, $\gamma = 0.95$, $\xi =0.001$ and $r_0 = 1.05$, respectively.}
\end{figure*}

\begin{figure*}
\centering \epsfig{file=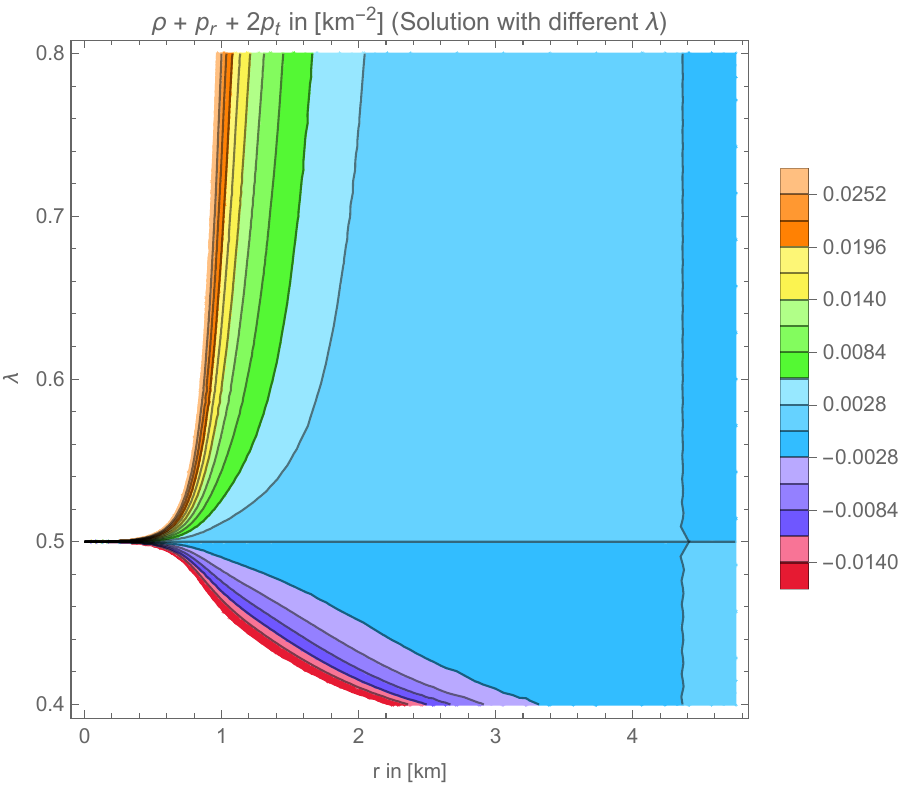, width=.35\linewidth,
height=2.1in}\centering \epsfig{file=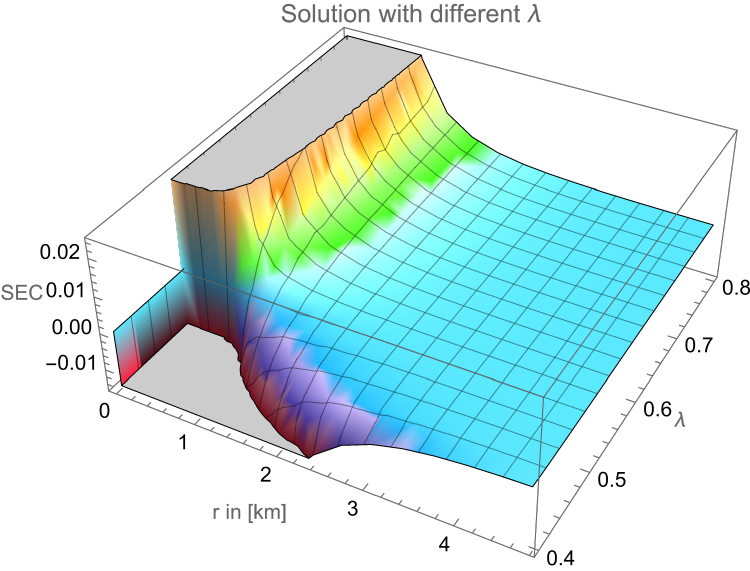, width=.35\linewidth,
height=2.1in}
\caption{\label{fig8} The graph consists of two panels: the left panel displays a contour plot, while the right panel presents a three-dimensional plot. It illustrates the behavior of the SEC i.e., ($\rho + p_r +2p_t$) for the wormhole solution with different values of $\lambda$ in the range of $\lambda \in [0.4, 0.8]$. The wormhole was characterized by specific parameter values, namely $\Sigma(x) = 0.95$, $\gamma = 0.95$, $\xi =0.001$ and $r_0 = 1.05$, respectively.}
\end{figure*}

Since, we know that WEC or SEC $\implies$ NEC and DEC $\implies$ WEC $\implies$ NEC. Interestingly, all  ECs are automatically violated if NEC is violated.  Thus, our main focus lies on the NEC. In Figs. 
\ref{fig3} to \ref{fig8}, we depict all ECs i.e.,  NEC, WEC, DEC, and the SEC along radial and tangential directions. Observing the Figs. \ref{fig3} and \ref{fig4}, we see that the inequality $\rho + p_r < 0$ holds at the throat of the wormhole and its vicinity only for $\lambda \leq 0.5$. Similarly, the inequality $\rho + p_t < 0$ also holds at the throat and its vicinity of the wormhole model for $\lambda \leq 0.5$ and $\lambda \geq 0.6$. From these observations, we can say that 
 the model violates NEC throughout the spacetime and consequently the violation of WEC also in regions where $\lambda \leq 0.5$ and $\lambda \geq 0.6$. Interestingly, the DEC is also violated at the wormhole's throat and its neighborhood, as evident from Figs. \ref{fig5}, \ref{fig6} and \ref{fig7}, respectively. On the other hand, the SEC remains positive for $\lambda \geq 0.5$, indicating compliance with the ECs, as shown in Fig. \ref{fig8}.

\section{Discussion and Implications}\label{Sec6}

The physical importance of Rastall-Rainbow gravity is profound regardless of any modified gravity theories. 
Since, the Rastall-Rainbow gravity model has been constructed with the combination of Rastall theory and the gravity's rainbow formalism. The present article aims to explore the possible existence of static and spherically symmetric wormholes in Rastall-Rainbow gravity. The modeling has been carried out using the well known embedding class I techniques to embed a four-dimensional space-time into a
pseudo-Riemannian space-time. This procedure helps us to establish a connection between two metric potentials, and thus
we are able to solve the modified field equations analytically. 

More specifically, we obtained exact wormhole solutions by considering a specific choice of redshift function and solve the field equations analytically. We apply stringent constraints to our model parameters ($\lambda, \Sigma[x]$) from the flaring-out conditions and investigate the effects of both parameters on it. Notably, the resulting shape function obtained from the embedding approach is depending on both model parameters $\lambda$ and  $\Sigma[x]$. Taking into account these constraints, we further investigate the energy conditions and showed that wormhole space-time
violates the NEC at the throat and its vicinity, and consequently the violation of WEC also. Additionally, we investigate the
overall stability of the configuration by determining the adiabatic sound velocity and explicitly computed the stable region, as shown in Fig. \ref{fig2}. Finally, we can say that the present approach may constitute a stable wormhole solution in Rastall-Rainbow gravity theory. \\
%-----------------
%-----------------

%\section*{Appendix}\label{Sec8}

\section*{Acknowledgments} 
This study is supported via funding from Prince Sattam bin Abdulaziz University project number (PSAU/2024/R/1445). The authors are thankful to the Deanship of Graduate Studies and Scientific Research at University of Bisha for supporting this work through the Fast-Track Research Support Program. AE thanks the National Research Foundation of South Africa for the award of a postdoctoral fellowship.

\section*{Conflict Of Interest statement }
The authors declare that they have no known competing financial interests or personal relationships that could have appeared to influence the work reported in this paper.

\section*{Data Availability Statement} 
This manuscript has no associated data, or the data will not be deposited. (There is no observational data related to this article. The necessary calculations and graphic discussion can be made available
on request.)



\end{document}